\documentclass[review,number,sort&compress]{elsarticle}

\usepackage{graphicx}
\usepackage{color}
\usepackage{calc}
\usepackage[utf8]{inputenc}
\usepackage{epsfig}
\usepackage{amsmath}
\usepackage{amsfonts}
\usepackage{soul}

\definecolor{ColorOfLink}{rgb}{0.5,0,0.5}
\definecolor{ColorOfFile}{rgb}{0,0.5,0}
\definecolor{ColorOfURL}{rgb}{0,0,0.7}

\def\EE{\mathcal{E}}

\def\CC{\mathfrak{C}}
\def\SS{\mathfrak{S}}

\def\FWHM{\Delta_{\mathrm{inst}}}

\journal{Nucl. Instr. and Meth. A}

\begin{document}
\begin{frontmatter}
\title{Title\tnoteref{lab:title}}
\tnotetext[label1]{}

\cortext[cor1]{Corresponding author}

\author[aff:1,aff:2]{G.~A.~Kazakov\corref{cor1}}
\ead{kazakov.george@gmail.com}

\author[aff:2]{V.~Schauer}
\author[aff:2]{J.~Schwestka}
\author[aff:2]{S.~P.~Stellmer}
\author[aff:2]{J.~H.~Sterba}
\author[aff:4]{A.~Fleischmann}
\author[aff:4]{L.~Gastaldo}
\author[aff:4]{A.~Pabinger}
\author[aff:4]{C. Enss}
\author[aff:2]{T.~Schumm}

\address[aff:1]{Wolfgang Pauli Institute, Univ. Wien - UZA 4 Nordbergstrasse 15, A 1090, Vienna, Austria}
\address[aff:2]{Vienna Center for Quantum Science and Technology, Atominstitut, TU Wien, Stadionallee 2, 1020, Vienna, Austria}
\address[aff:4]{Kirchhoff-Institute for Physics, Heidelberg University, INF 227, 69120 Heidelberg, Germany}

\title{Prospects for measuring the $^{229}$Th isomer energy using a metallic magnetic microcalorimeter}


\author{}

\address{}


\begin{abstract}
The Thorium-229 isotope features a nuclear isomer state with an extremely low energy. The currently most accepted energy value, $7.8 \pm 0.5$\,eV, was obtained from an indirect measurement using a NASA x-ray microcalorimeter with an instrumental resolution 26\,eV. We study, how state-of-the-art magnetic metallic microcalorimeters with an energy resolution down to a few eV can be used to measure the isomer energy. In particular, resolving the 29.18\,keV doublet in the $\gamma$-spectrum following the $\alpha$-decay of Uranium-233, corresponding to the decay into the ground and isomer state, allows to measure the isomer transition energy without additional theoretical input parameters, and increase the energy accuracy. We study the possibility of resolving the 29.18\,keV line as a doublet and the dependence of the attainable precision of the energy measurement on the signal and background count rates and the instrumental resolution.
\end{abstract}

\end{frontmatter}
\newpage
\section{Introduction}
\label{sec:intro}

The nuclear level scheme of the Thorium-229 isotope is expected to feature a long-lived isomer state, $^{229m}$Th, extremely close to the nuclear ground state. The most recent value for the isomer energy $E_{\mathrm{is}}$, $7.8 \pm 0.5$\,eV, obtained from indirect measurements with a NASA  x-ray microcalorimeter\footnote{We will refer to these devices as ``x-ray'' spectrometers, corresponding to their primary field of application. In the measurements described here, they detect both, x-rays and $\gamma$-rays.}~\cite{Beck07, Beck09}, is within the reach of modern optical laser spectroscopy and could serve as a ``nuclear frequency standard''~\cite{Peik03}. This standard could reach an uncertainty level of  $10^{-19}$~\cite{Campbell11}, and provide a new powerful instrument for testing the stability of fundamental constants~\cite{Litvinova09, BerengutIsomer09}. It has been shown that an ensemble of Thorium nuclei doped into a transparent crystal may demonstrate superradiance with a non-trivial emission dynamics~\cite{Liao12}, and may be used for building an ultraviolet (UV) laser~\cite{Tkalya11}. Finally, the frequency shifts and broadenings produced by such a crystal lattice environment might be used in studies of material properties, as is commonly done in M\"ossbauer spectroscopy~\cite{Peik03}. The necessary step towards all of these exciting applications is a direct observation and precise determination of the isomer state energy.

\begin{figure}
\begin{center}
\resizebox{0.98\textwidth}{!}{\includegraphics{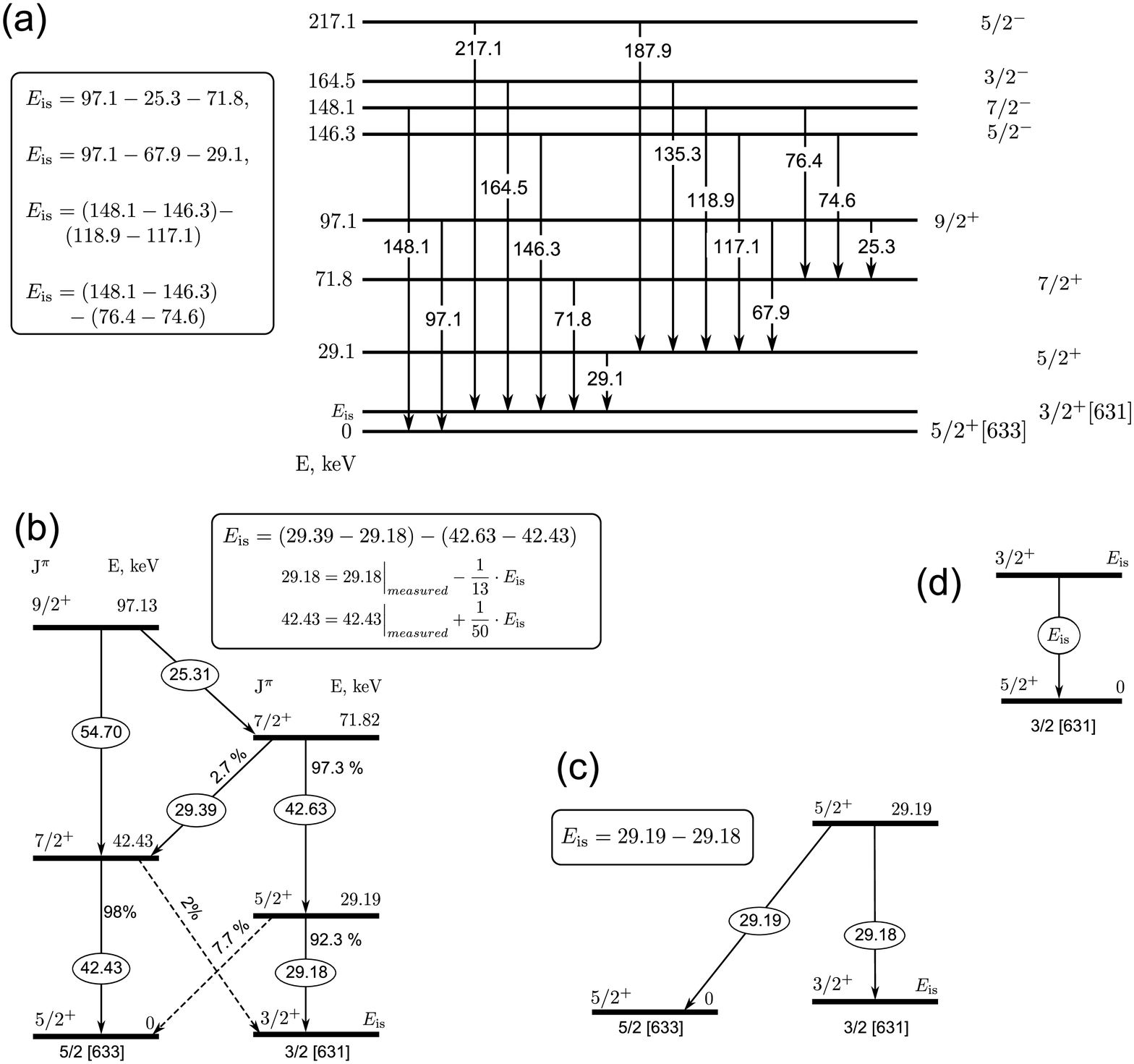}}
\end{center}
\caption{Partial level schemes of the $^{229}$Th nucleus with decay paths and energies (all in keV). Boxes in each panel denote the energy combinations used to derive $E_{\mathrm{is}}$ in the ``indirect'' methods discussed in the main text. (a): according to Helmer and Reich \cite{Helmer94}; (b): according to Beck et al.~\cite{Beck07, Beck09}, the interband transitions (dashed arrows) are taken into account; (c): approach discussed here using a high-resolution ($\FWHM \simeq$ 3 -- 9\,eV) microcalorimeter to resolve the 29.1\,keV doublet (first proposed in \cite{BeckTrento}); (d): direct detection of ``nuclear light'' (many unsuccessful attempts~\cite{Irwin97, Richardson98, Shaw99, Utter99} and new proposal~\cite{Wense13}). Schemes (a), (b), and (c) are {\em indirect} measurements, involving keV energy transitions whereas scheme (d) is {\em direct}, only measuring the isomer transition of a few eV energy.}
\label{fig:f1}
\end{figure}

The existence of the low-energy state in the $^{229}$Th nucleus was first conjectured by Kroger and Reich based on studies of the $\gamma$-ray spectrum following the $\alpha$-decay of Uranium-233~\cite{Kroger76}. They concluded that this nucleus has a $J^{\pi}=3/2^+$ isomer level lying within $100$\,eV above the $J^{\pi}=5/2^+$ ground state level. The development of high quality germanium detectors (resolution from 300 to 900\,eV) allowed Helmer and Reich to measure more precise $\gamma$-energies in 1989 -- 1993 and to predict the energy of the nuclear transition to be $E_{\mathrm{is}}=3.5 \pm 1.0$\,eV, placing it into the range of optical frequencies~\cite{Helmer94}. The decay pattern and combinations of transitions used by Helmer and Reich are presented in Figure \ref{fig:f1} (a).

This unnaturally low value of $E_{\mathrm{is}}$ triggered a multitude of investigations, both theoretical and experimental, trying to determine the transition energy precisely, and to specify other properties of the $J^{\pi}=3/2^+$ excited state (such as lifetime and magnetic moment). However, searches for  direct photon emission from the low-lying excited state performed in the late 90's \cite{Irwin97, Richardson98} have failed to observe a signal~\cite{Shaw99, Utter99}. In 2005, Guim\~araes-Filho and Helene re-analysed the data of Helmer and Reich, taking into account new information about the nuclear decay pattern~\cite{Filho05}. They derived $E_{\mathrm{is}}=5.5 \pm 1.0$\,eV. 

In 2007, a cryogenic NASA x-ray microcalorimeter with instrumental resolution $\FWHM$ from 26 to 30\,eV (FWHM) allowed Beck et al. \cite{Beck07} to perform a new indirect measurement of $E_{\mathrm{is}}$, involving lower energy nuclear states, as depicted in Figure \ref{fig:f1} (b). In this measurement, the obtained transition energy (7\,eV) was corrected by accounting for the theoretical branching ratios $29.19 \,\mathrm{keV} \rightarrow\,^{229g}\mathrm{Th}$ estimated as 1/13, and $42.43 \, \mathrm{keV} \rightarrow \, ^{229m}\mathrm{Th}$ estimated as 2\% in~\cite{Beck09}. This correction yields the currently most accepted value $E_{\mathrm{is}}=7.8 \pm 0.5$\,eV, now placing the transition into the vacuum UV range ($\approx 160$\,nm).

In the experiments decribed above (\cite{Beck07, Kroger76, Helmer94}) the isomer transition energy $E_{\mathrm{is}}$ is not measured directly but is derived from the spectrum of higher-energy (keV) $\gamma$-radiation of a spontaneously decaying $^{233}$U source. We will refer to these measurements as {\em indirect passive}. Possible alternatives are {\em direct passive} and {\em active} approaches. 

In the {\em direct passive} schemes (Figure~\ref{fig:f1} (d)), the aim is to perform spectroscopy of the ultraviolet radiation emitted from the isomer appearing in the $\alpha$-decay of $^{233}$U (2\% of the nuclei decay is expected to lead into the isomer state). This method has two main difficulties: a relatively high false count rate caused by the Uranium sample radioactivity, and a high probability of non-radiative decay (quenching) of the isomer state in neutral Thorium atoms (up to $10^{9}$ times higher than the radiative decay rate \cite{Karpeshin07}). To overcome these problems, it was proposed in~\cite{Wense13} to extract $\alpha$-recoil Thorium ions ejected from an Uranium sample, and collect them in a small spot on a MgF$_2$ coated surface to minimize the quenching rate. Vacuum ultraviolet spectroscopy of the emitted fluorescence radiation may then allow to measure the isomer transition energy.

On the contrary, in {\em active approaches}, Thorium nuclei (in the ground state) will be illuminated by tunable radiation to excite them to the isomer state. In the {\em solid-state approach} a macroscopic ($10^{12}-10^{18}$) number of Thorium ions doped into UV transparent crystals can be excited, for example, by synchrotron radiation, and the emerging fluorescence signal can be studied~\cite{Rellergert10, We12, Hudson13}. Apparent advantage of this approach is the huge number of simultaneously excited nuclei. At the same time, crystal fluorescence can cause difficulties in identifying the Thorium isomer transition, and various crystal effects can hamper the precise determination of $E_{\mathrm{is}}$. Another approach is the {\em spectroscopy of trapped Thorium ions}. At PTB, Germany, work is under way to excite nuclei of Th$^{+}$ ions into the isomer state using a two-photon scheme, exploiting the electronic bridge mechanism~\cite{PorsevPeik10, Herrera12}. In Georgia University of Technology, USA, the laser manipulation of Th$^{3+}$ ions is under investigation ~\cite{Campbell11, Radnaev12}. Detection of the excitation of the Thorium into the isomer state may be based on a change of the electronic hyperfine structure~\cite{Peik03}. Studies of the hyperfine structure of Thorium are also performed at the IGISOL facility in Jyv\"askyl\"a, Finland, in collaboration with a group of the University of Mainz, Germany~\cite{Sonnenschein12}.

We should also mention a number of studies aimed to measure the lifetime of the isomer state without a determination of $E_{\mathrm{is}}$. In~\cite{Ruchowska06}, the half life of the isomer state for a \emph{bare nucleus} was derived theoretically based on the calculations of the matrix element of the nuclear magnetic moment and on the experimental data concerning transitions at higher energies. They predict a half-life of $T_{1/2}=(10.95\,\mathrm{h})/ (0.025 E^3)$ for the isomer transition, where $E$ is given in eV, which yields $T_{1/2}=55$\,min  for $E=7.8$\,eV. Direct measurements of this lifetime were performed in several groups~\cite{Inamura09, Kikunaga09, Zhao12}. The obtained results vary from $2$ min~\cite{Inamura09} to 6 hours~\cite{Zhao12}. This discrepancy may be explained either by an incorrect interpretation of the observations \cite{Peik13} or by a difference in chemical composition of the Thorium resulting in different internal conversion rates.

All active approaches and eventually all nuclear spectroscopy applications require irradiation of the sample with some external narrow-band tunable radiation, and study of the emerging fluorescence. The estimation of the error $\sigma=0.5$\,eV on the isomer energy presented in~\cite{Beck07} corresponds to one standard deviation, therefore it is necessary to scan the excitation source over 2\,eV ($\pm 2\sigma$) to find the transition with 95\,\% probability, or over 3\,eV ($\pm 3\sigma$) to find the transition with 99.7\,\% probability. Sakharov re-estimated the influence of the uncertainty of the 29.39\,keV peak on the isomer energy derivation in~\cite{Beck07} and obtained an error of 1.3--1.5\,eV~\cite{Sakharov10}. Moreover, an analysis of more recent experimental data led him to claim that the energy of the isomer state can be anywhere in the range 0--15\,eV, if the isomer state exists at all.

We believe it will be technically difficult, if not impossible, to cover such a broad energy range with a tunable narrow-band source of ultraviolet radiation in a reasonable time. We therefore propose to first increase the energy resolution on $E_{\mathrm{is}}$ by an improved indirect measurement compared to~\cite{Beck07}. As we show below, it appears possible to resolve the 29.18\,keV doublet~\cite{BeckTrento} presented in Figure~\ref{fig:f1} (c) with todays state-of-the-art x-ray spectrometers. Resolving this doublet would significantly increase confidence in the existence of the isomer state. Moreover, the isomer energy would be measured without additional theoretical input parameters like branching ratios etc. The aim of the present study is to investigate the possibility of resolving the 29.18\,keV line clearly as a doublet over a broad range of values for the isomer energy splitting and the branching ratio, and to analyze the precision that can be obtained on $E_{\mathrm{is}}$ depending on the relevant experimental parameters.

\section{Statistical aspects of the envisioned experiment}

The operation principle of high-resolution x-ray microcalorimeters is to detect the heat deposited by an x- or $\gamma$-ray interacting with an absorber, using a very sensitive thermometer. Interaction with the absorber material mainly proceeds through the photoeffect. The energy of the produced photoelectron as well as Auger electrons together with their thermalization cascade should be effectively deposited within the volume of the absorber~\cite{Porter04, Pies12}. On the other hand, the absorber should have a small heat capacity $C_a$ for good instrumental energy resolution $\FWHM$. Various microcalorimeters differ in geometry, absorber material, sensor, etc., which leads to different energy resolutions, stopping powers, total detector surfaces etc. Many of these parameters are connected and can not be optimized independently. For example, increasing the size of absorber increases the solid angle and/or stopping power but degrades the instrumental energy resolution. Finally, we note that after a detection event, dissipation of the deposited heat leads to a detector-specific dead time, during which the energy of a successive photon can not be measured correctly. Therefore it is impossible to improve the precision of the measurements infinitely simply by using a more active sample, or by placing the sample very close to the detector. The total count rate can be reduced using a designed filter which will primarily absorb photons outside the 29.18\,keV region of interest. 

The present study aims to answer two questions: how does the possibility to resolve the 29.18\,keV peak as a doublet depend on the experimental parameters, and how to attain the most precise determination of the isomer transition energy $E_{\mathrm{is}}$. As outlined above, parameters of the experimental setup can be controlled to some extent only. For a proper design of suitable detectors and experimental configurations, we analyze how the key parameters affect these two points.

\subsection{Specification of the problem and statistical model}
The model employed for the statistical study should not contain too many parameters to make it accessible to a multi-factor analysis. On the other hand, it should be sufficiently comprehensive for a realistic feasibility study. For the sake of convenience, we assume a fixed total measurement time of $t=10^6$\,s, approximately 11 days, which corresponds to the total time of the successfull measurement in~\cite{Beck07}. Also we suppose that the background count rate near the 29.18\,keV doublet is flat and symmetric, and that the monoenergetic line has Gaussian shape with full width at half maximum equal to $\FWHM$ (see section~\ref{sec:issues} for a discussion of these approximations). 

The considered total energy interval (0-70\,eV) is subdivided to a set of 0.4\,eV bins (approximately a factor 10 below the expected instrumental energy resolution). The number of counts in the $i$th energy bin is a Poissonian random number $n_i$ with a mean value $\lambda_i$ equal to
\begin{equation}
\begin{split}
\lambda_{i}=&\frac{d \cdot R_{29} \cdot t}{\sigma \sqrt{2 \pi}}
\left[(1-b)\exp \left(-\frac{(E_i-\EE_1)^2}{2 \sigma^2}\right) + \right. \\
&\left. b \exp \left(-\frac{(E_i-\EE_2)^2}{2 \sigma^2} \right)
\right]+d\cdot r_{\mathrm{bg}}\cdot t, \label{ee:1}
\end{split}
\end{equation}
where $R_{29}$ is the signal count rate, $r_{\mathrm{bg}}$ is a specific rate of background counts per 1\,eV energy interval, $\EE_1$, $\EE_2$ are the centers of lines of the components of the 29.18\,keV doublet, $\sigma= \frac{\FWHM}{2\sqrt{2 \mathrm{ln}(2)}}$, $d=0.4$\,eV is one energy bin, $E_i$ is the mean energy of the $i$th bin, and $b$ is the branching ratio. The set $\{n_1,...,n_N\}$ of experimental data can be represented as a position vector $\mathbf{n}$ in an $N$-dimensional ``sample space'' ($N=175$).

For a given sample $\mathbf{n}$, we perform a nonlinear regression fit by a vector function $\mathbf{f}$ with $N$ components
\begin{equation}
\begin{split}
f_i=&\frac{d \cdot t}{s \sqrt{2 \pi}}\left[
J_1 \exp \left(-\frac{(E_i-\tilde{\EE_1})^2}{2 s^2}\right) + \right. \\
& \left. J_2 \exp \left(-\frac{(E_i-\tilde{\EE_1}-\tilde{E}_{is})^2}{2 \tilde{\sigma}^2} \right)
\right]+d\cdot \tilde{r}_{bg}\cdot t. \label{ee:2}
\end{split}
\end{equation}
This fit has 6 free parameters $\{J_1, J_2, \tilde{\EE_1},\tilde{E}_{is}, \tilde{r}_{bg}, \tilde{\sigma}\}=\{\theta_1,...,\theta_6\}\equiv \theta$. For the estimation of these parameters, we use the maximum likelihood method. The likelihood function $L(\mathbf{n}|\,\theta)$ is the probability for realizing the set $\mathbf{n}$, if true mean values $\lambda_i$ are equal to $f_i(\theta)$. We also introduce the logarithmic likelihood function
\begin{equation}
\ell(\mathbf{n}|\,\theta)=\log L(\mathbf{n}|\,\theta)=\sum_{i=1}^N \left[n_i\log f_i(\theta)-\log (n_i!)-f_i(\theta) \right]. \label{ee:3}
\end{equation}
%
\subsection{Resolving the 29.18\,keV line as a doublet}
As it was outlined above, resolving the 29.18\,keV line as a doublet would significantly reduce the doubts~\cite{Sakharov10} in the existence of the isomer state in the $^{229}$Th nucleus. Our first aim is to discuss the feasability of such an identification depending on experimental parametes and the (yet unknown) values of the isomer energy splitting $E_{\mathrm{is}}=\EE_2-\EE_1$ and the branching ratio $b$. 
\begin{figure}
\begin{center}
\resizebox{0.45\textwidth}{!}{\includegraphics{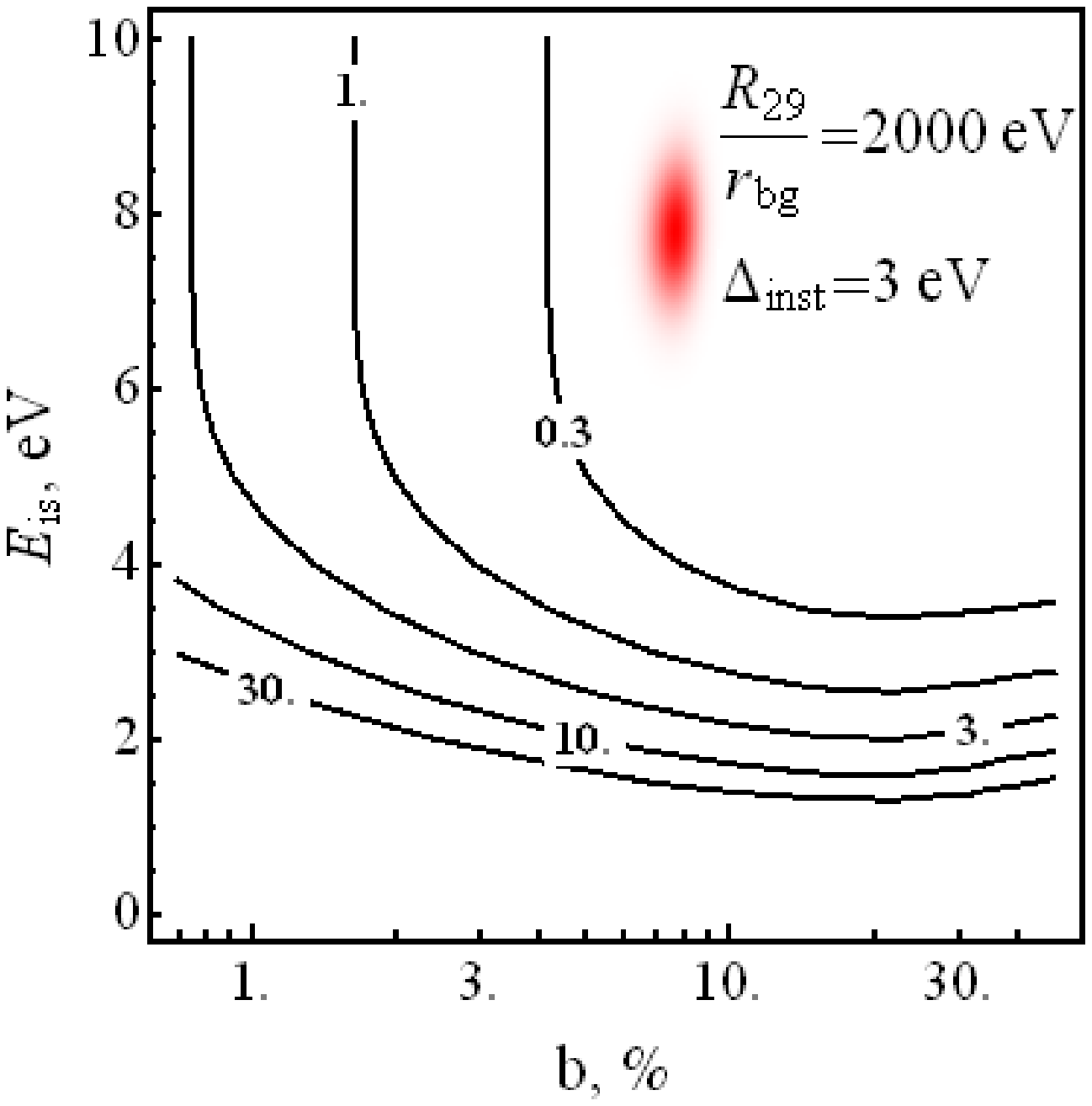}} \hspace{0.02\textwidth} 
\resizebox{0.45\textwidth}{!}{\includegraphics{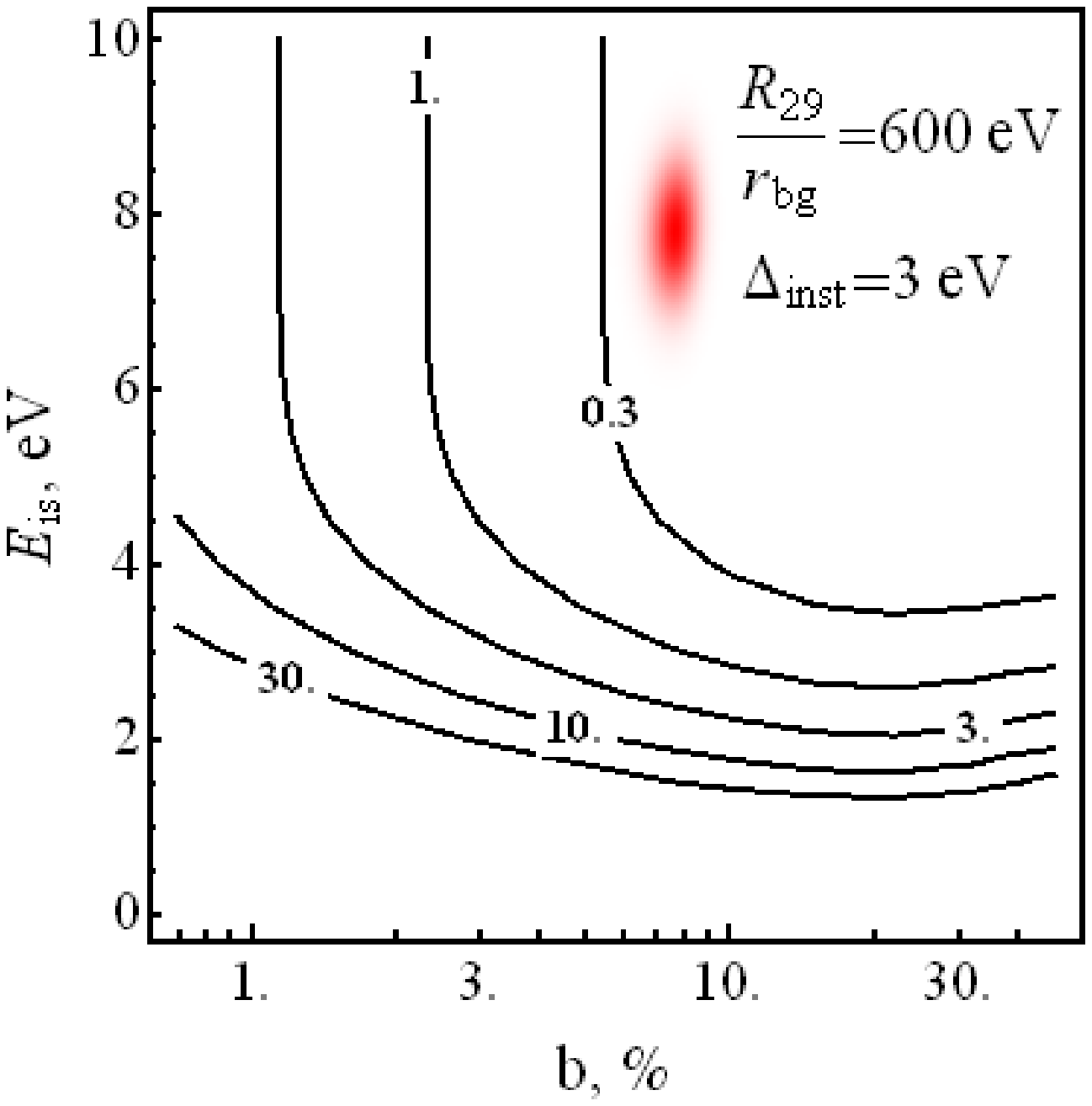}}\\
\resizebox{0.45\textwidth}{!}{\includegraphics{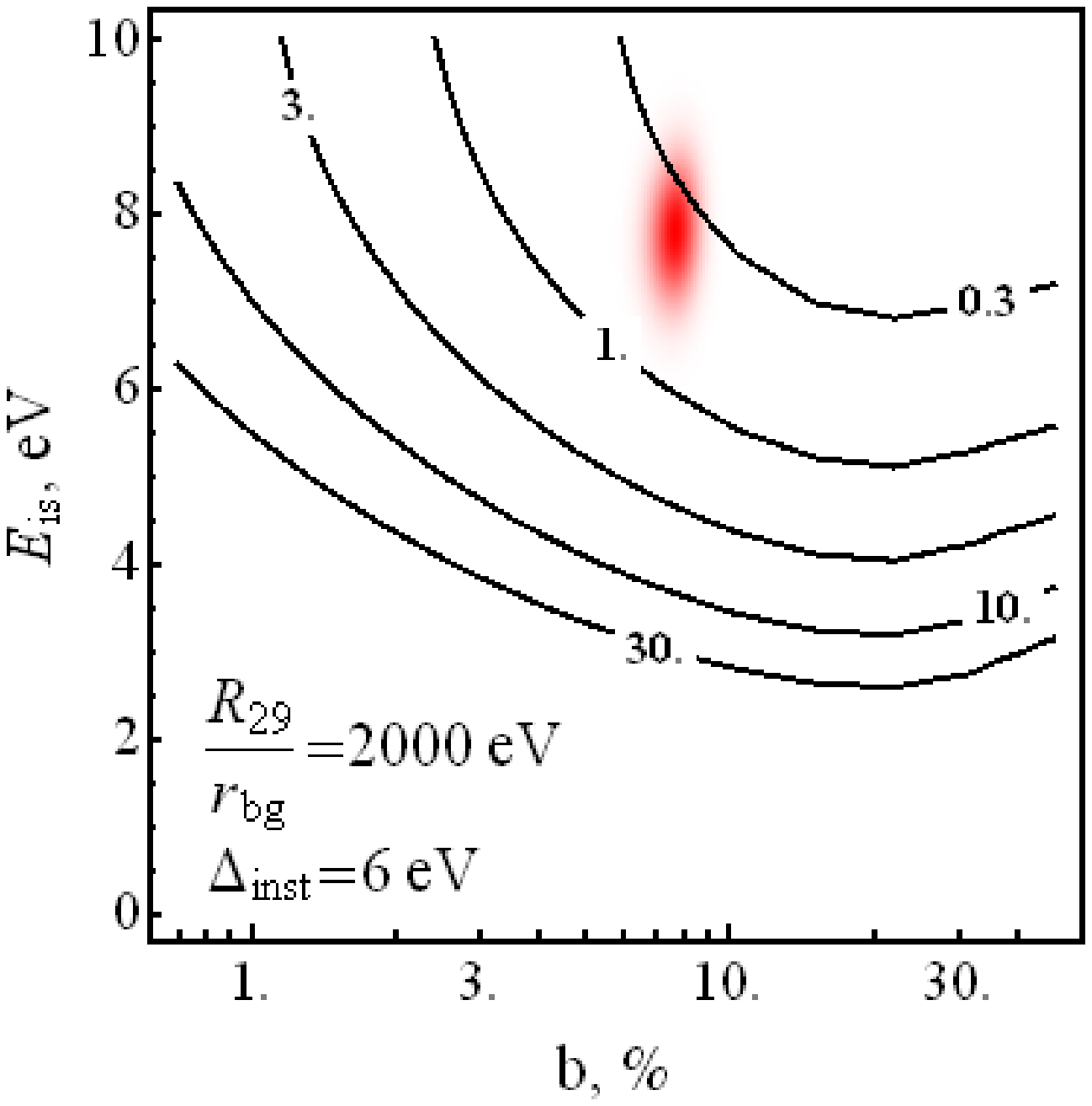}} \hspace{0.02\textwidth} 
\resizebox{0.45\textwidth}{!}{\includegraphics{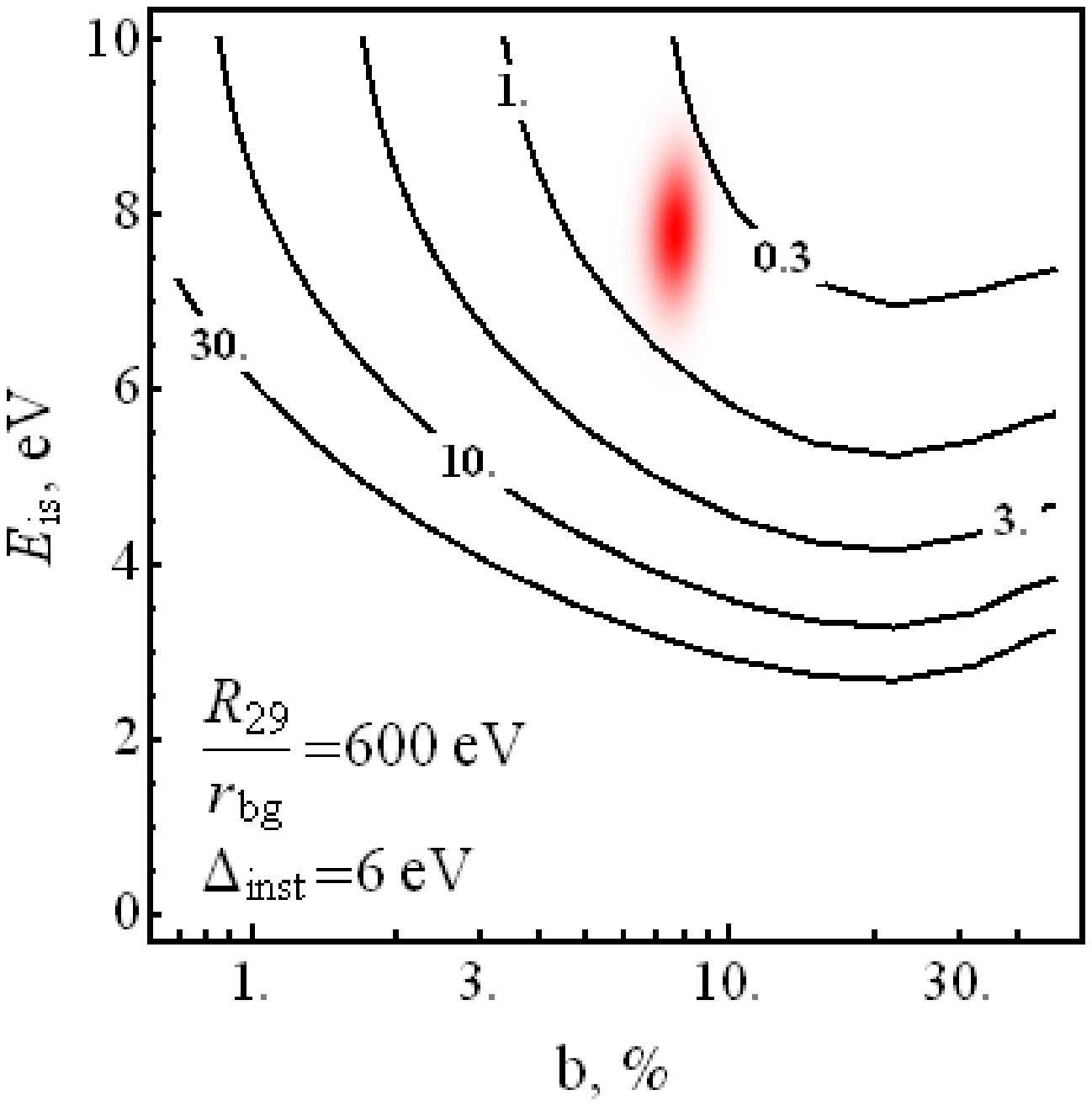}}\\
\resizebox{0.45\textwidth}{!}{\includegraphics{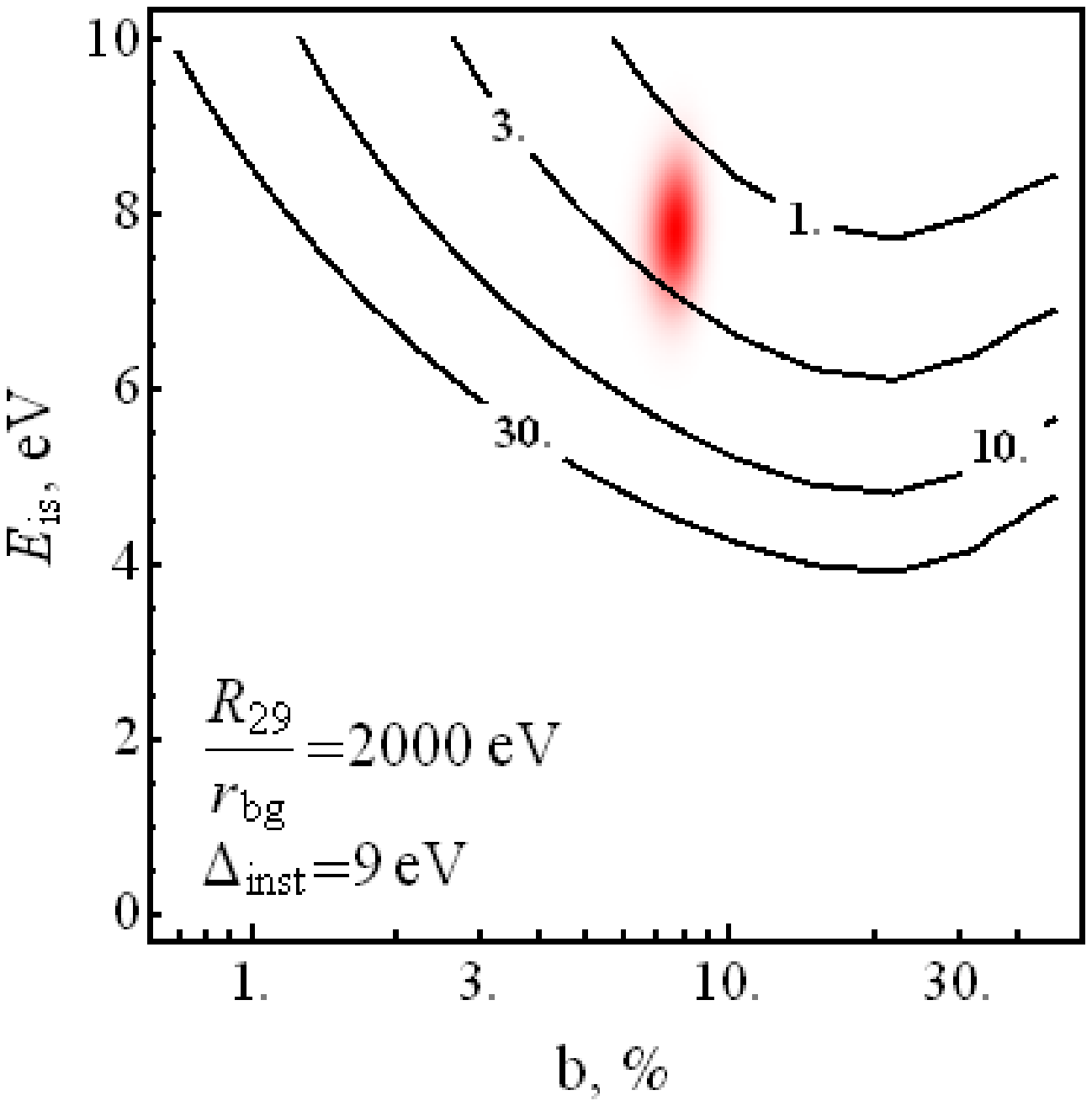}} \hspace{0.02\textwidth} 
\resizebox{0.45\textwidth}{!}{\includegraphics{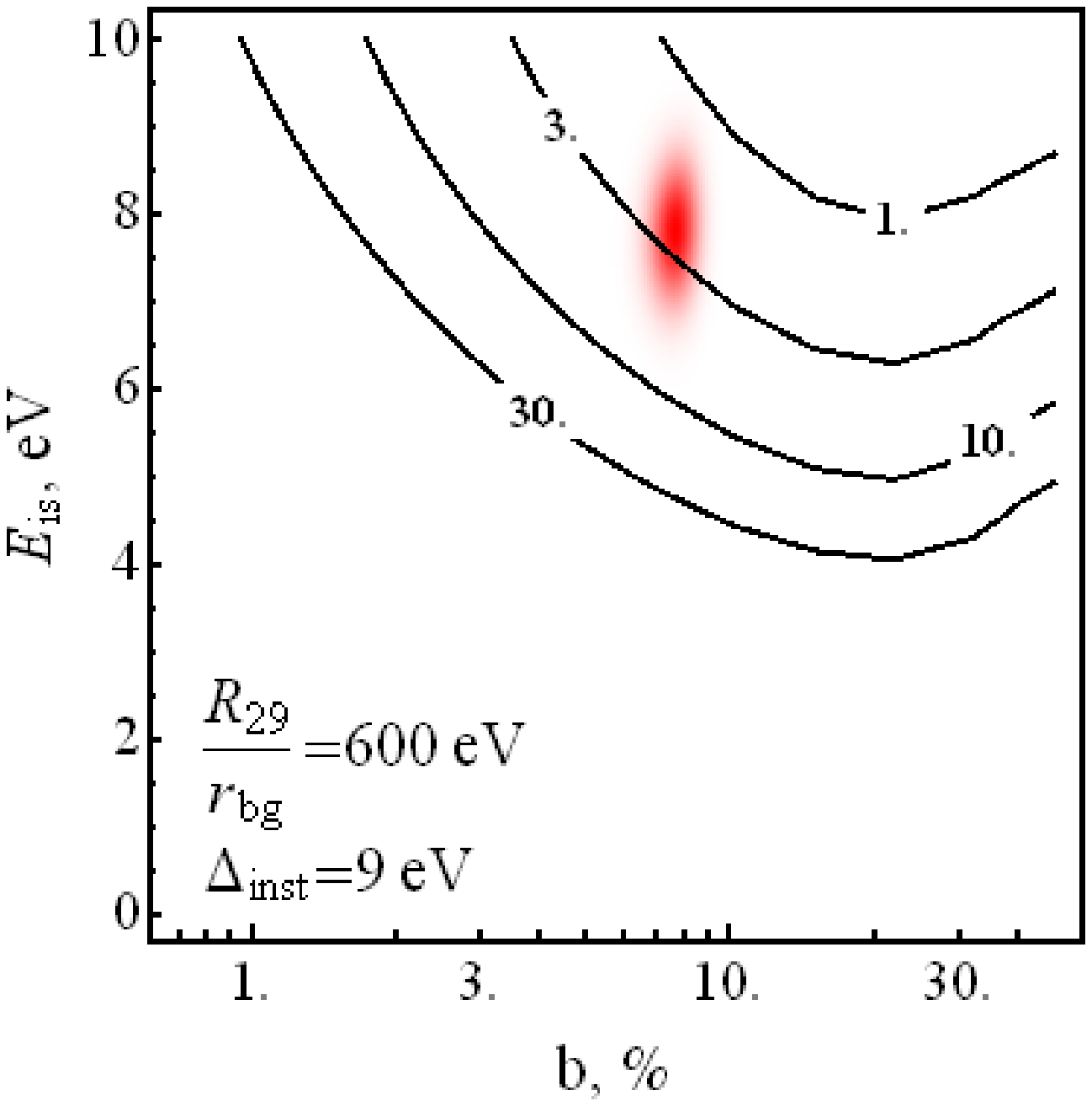}}
\end{center}
\caption{Curves of constant levels of signal count rate $R_{29}$ (in mHz) required to resolve the 29.18\,keV line as a doublet at 1\,\% significance level for different values of the detector resolution $\FWHM$ and the signal-to-noise ratio $R_{29}/r_{\mathrm{bg}}$ at $10^6$ seconds of measurement time. The red spot corresponds to the area of the branching ratio $b$ and isomer transition energy $E_{\mathrm{is}}$ according to \cite{Beck09}. }
\label{ff:2}
\end{figure}
\begin{figure}
\begin{center}
\resizebox{0.45\textwidth}{!}{\includegraphics{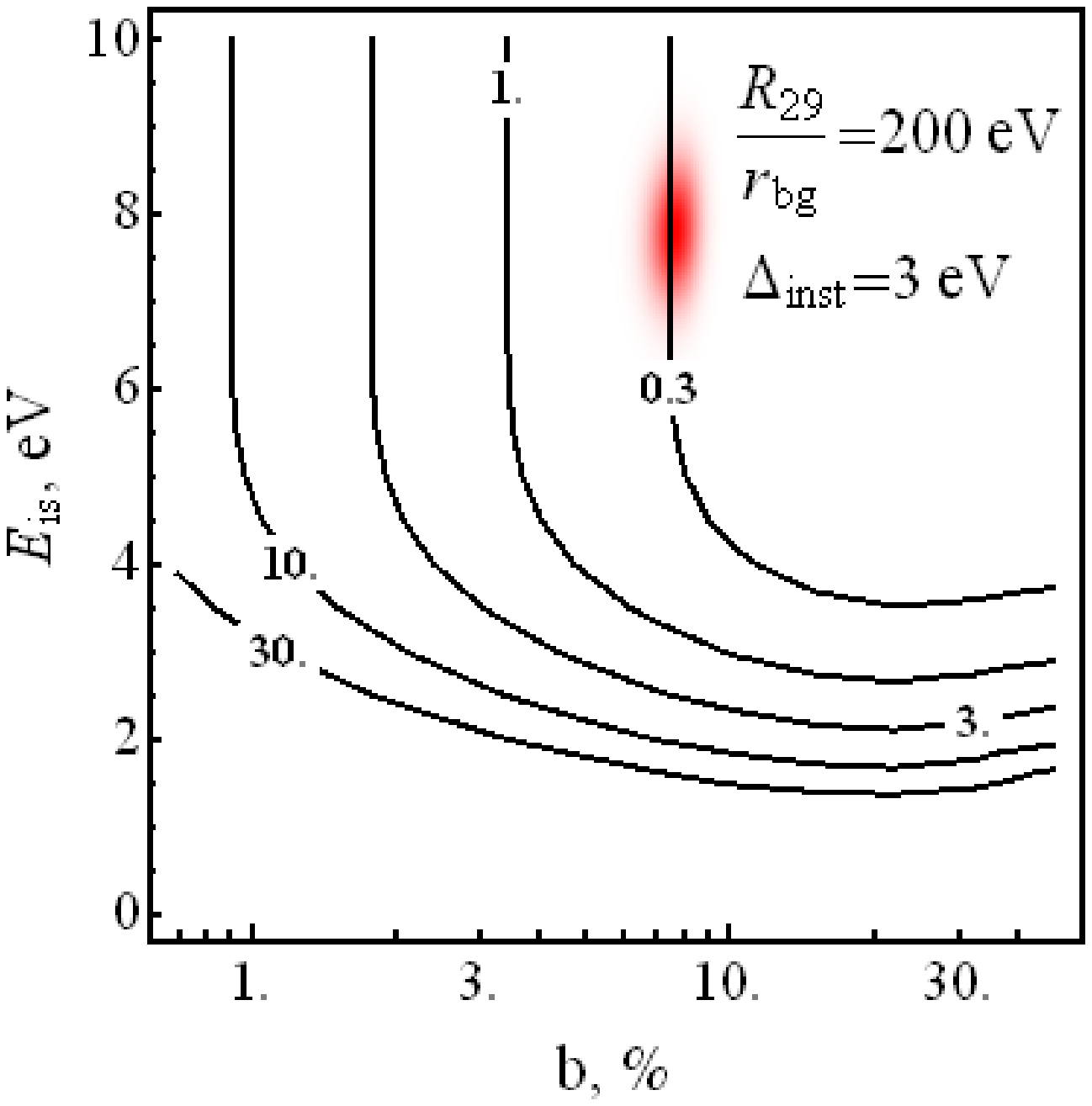}} \hspace{0.02\textwidth} 
\resizebox{0.45\textwidth}{!}{\includegraphics{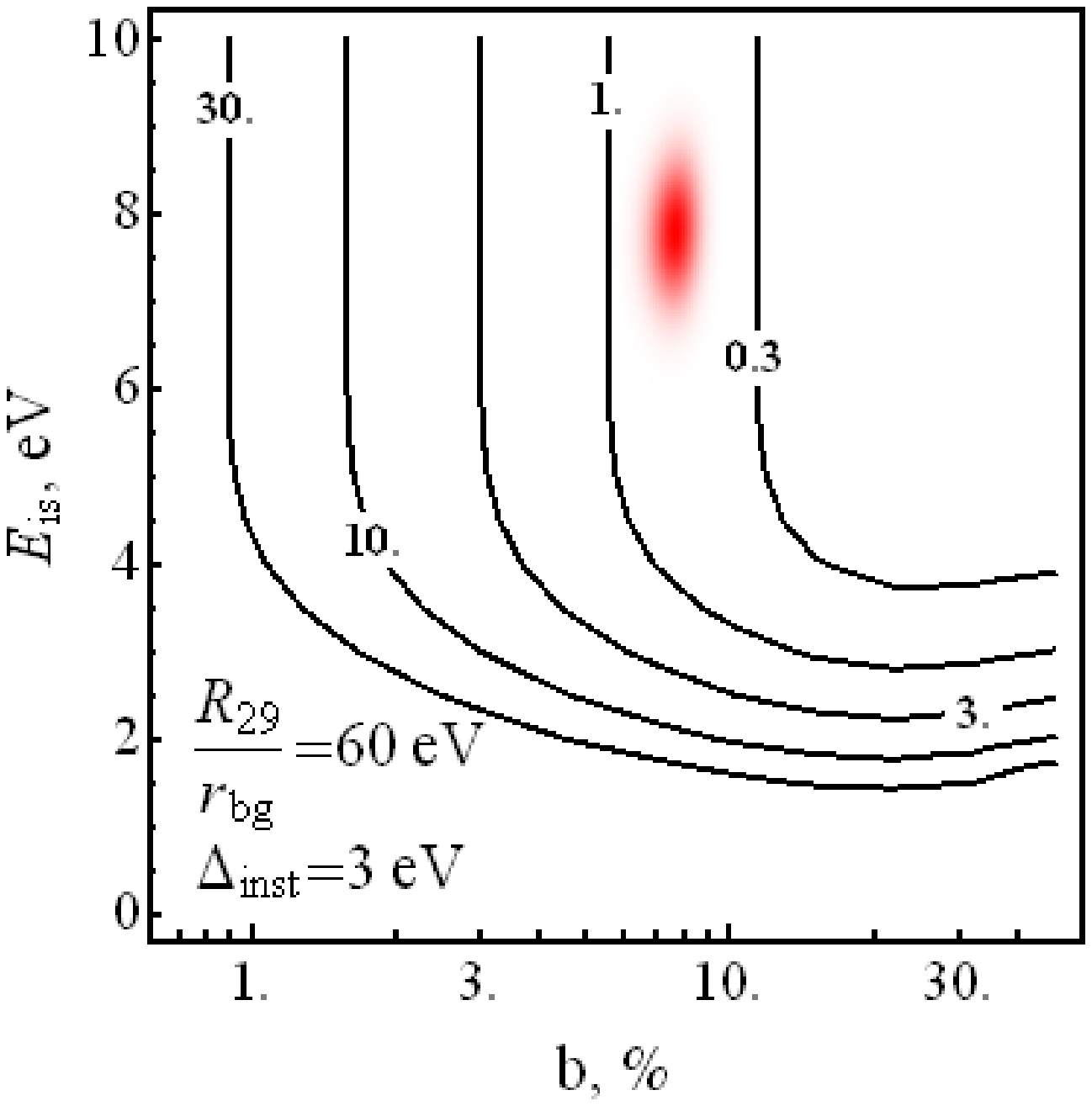}}\\
\resizebox{0.45\textwidth}{!}{\includegraphics{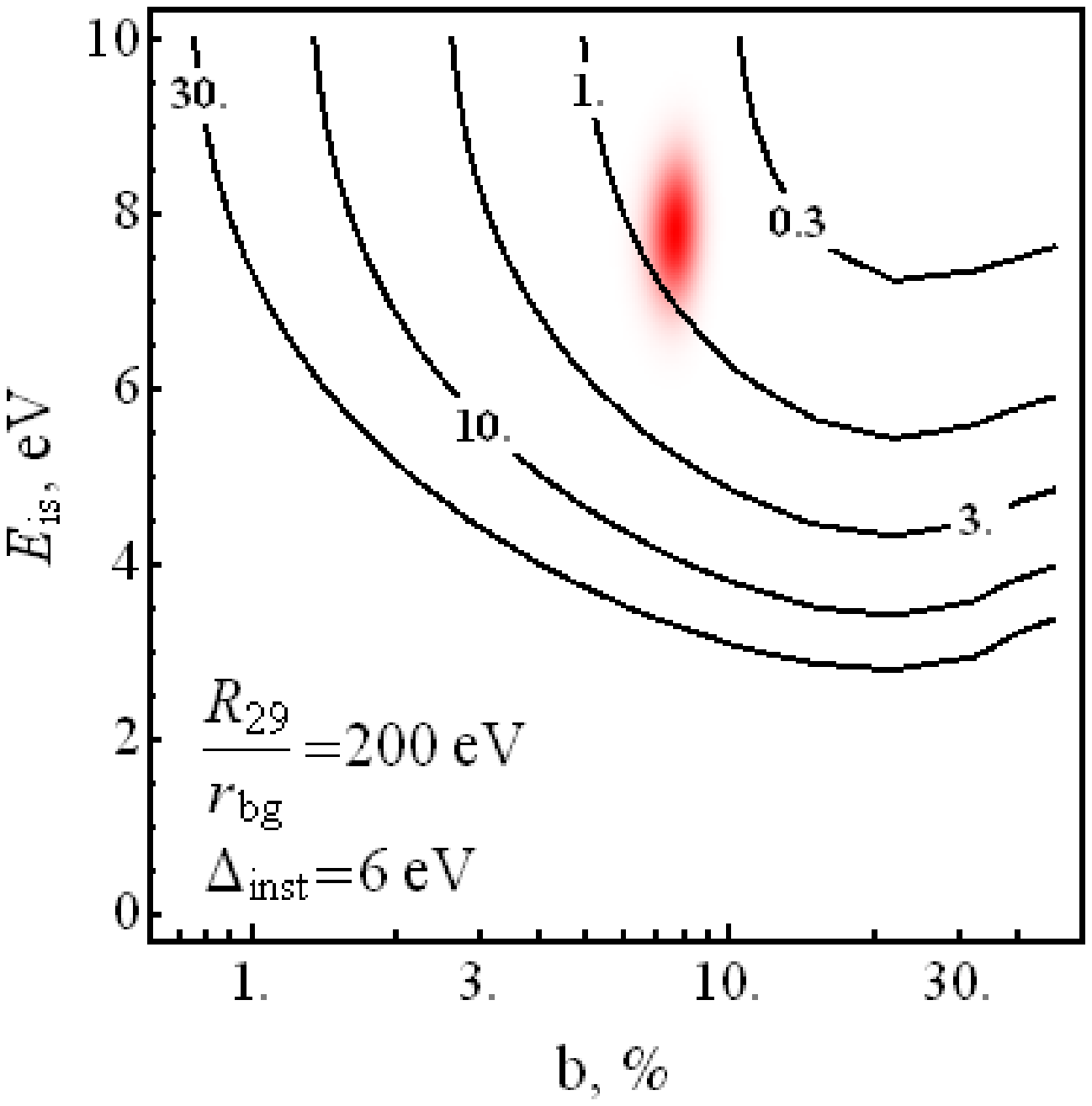}} \hspace{0.02\textwidth} 
\resizebox{0.45\textwidth}{!}{\includegraphics{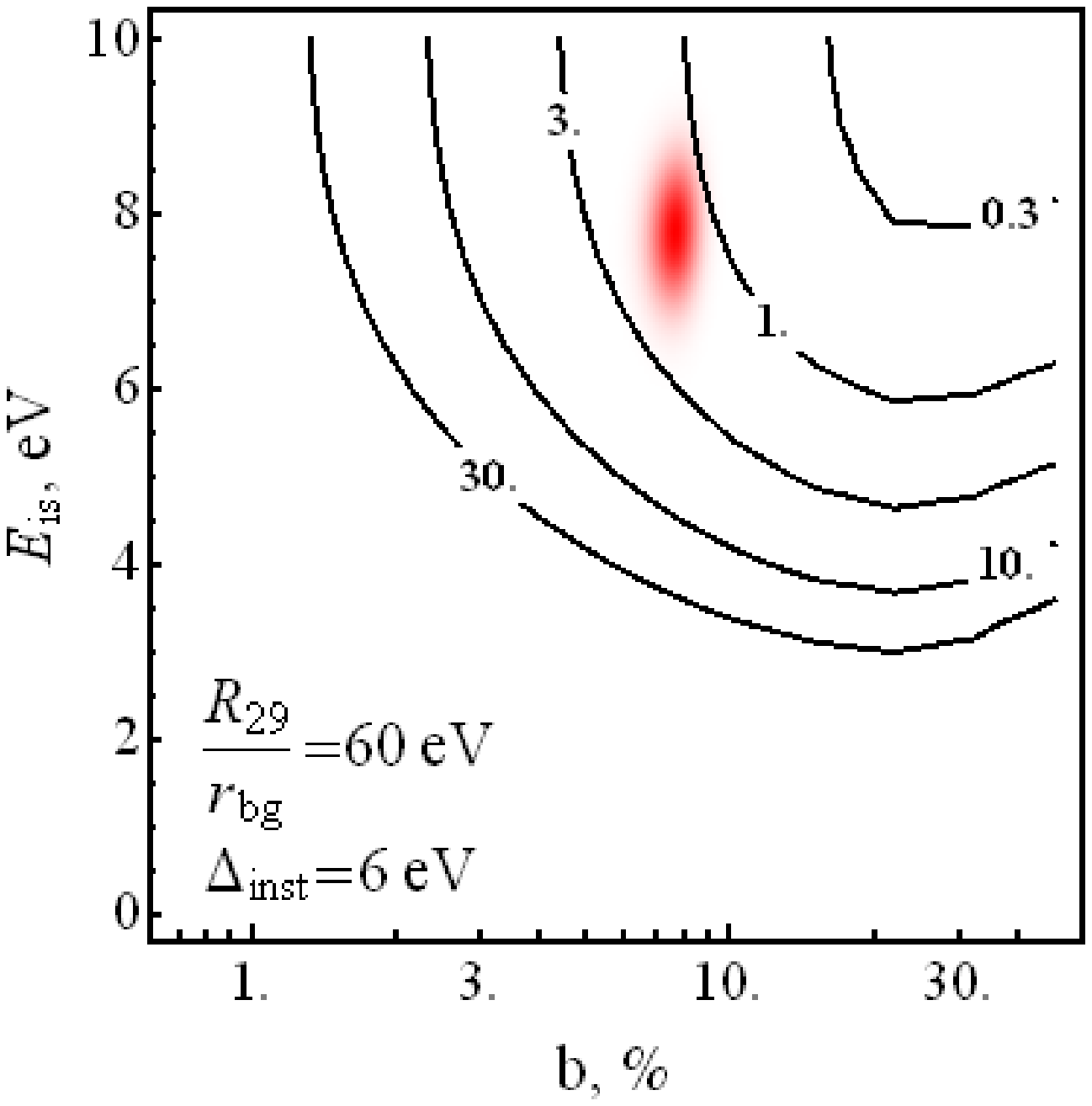}}\\
\resizebox{0.45\textwidth}{!}{\includegraphics{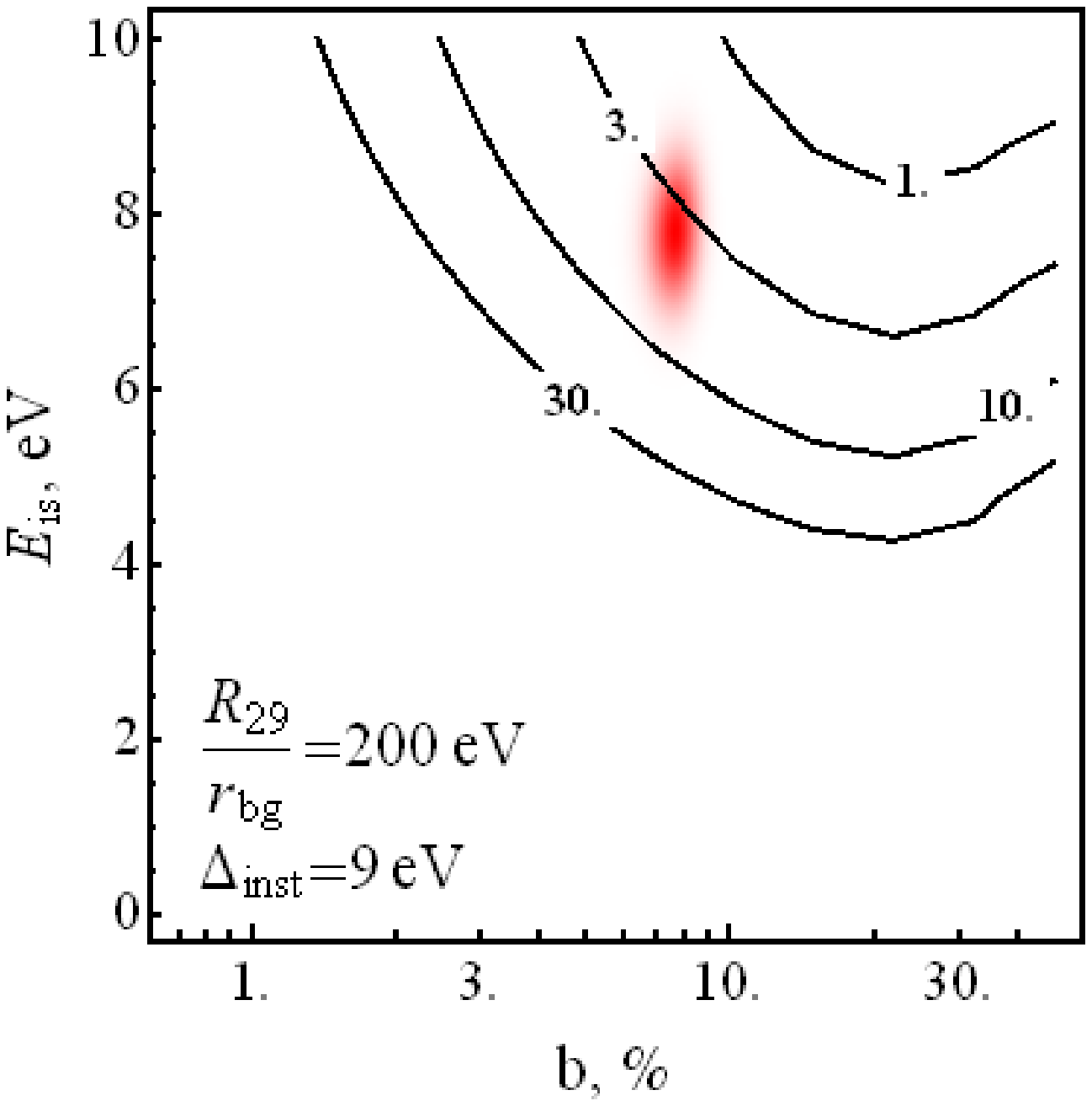}} \hspace{0.02\textwidth} 
\resizebox{0.45\textwidth}{!}{\includegraphics{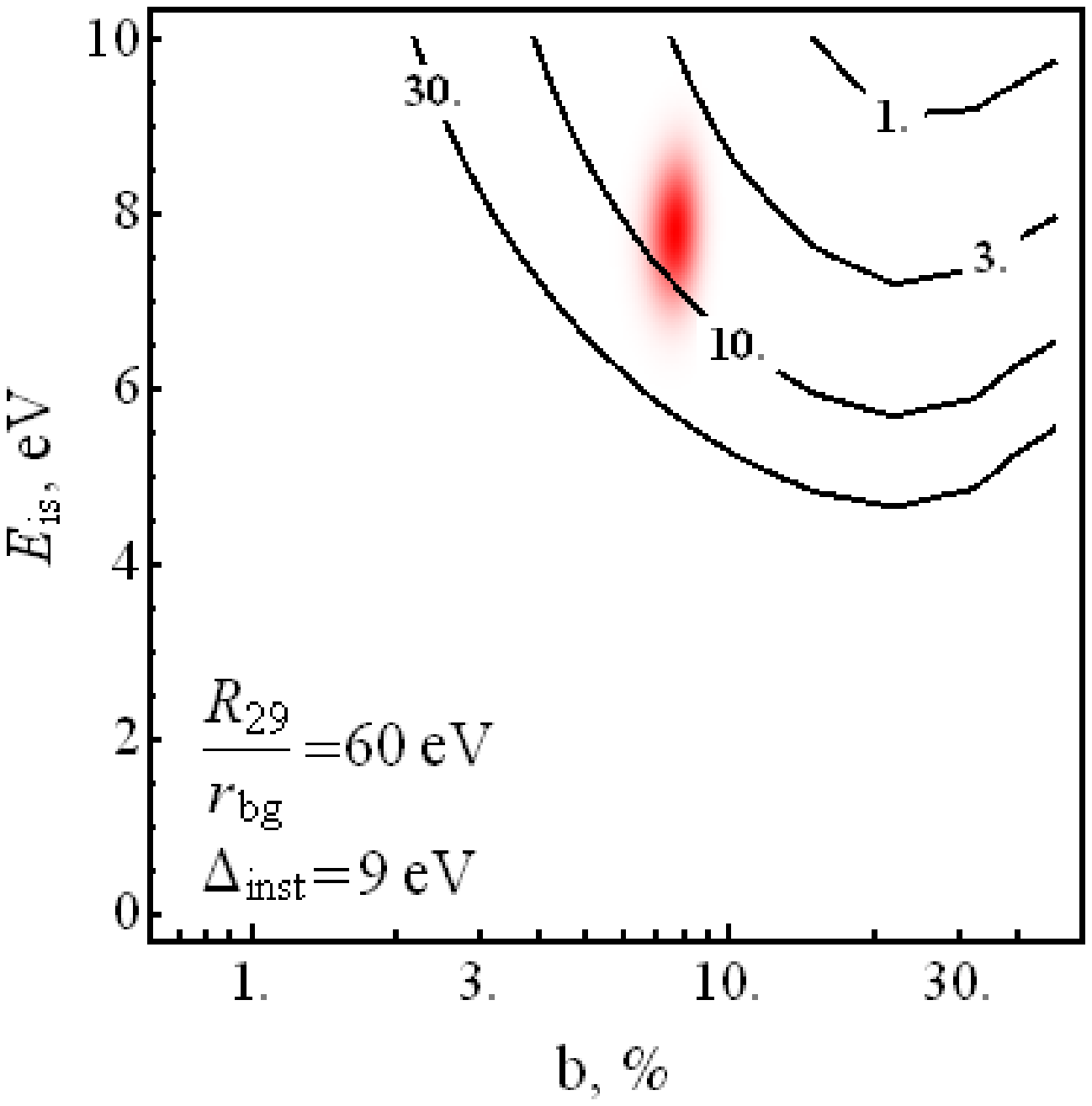}}
\end{center}
\caption{Curves of constant levels of signal count rate $R_{29}$ (in mHz) required to resolve the 29.18\,keV line as a doublet at 1\,\% significance level for different values of the detector resolution $\FWHM$ and the signal-to-noise ratio $R_{29}/r_{\mathrm{bg}}$ at $10^6$ seconds of measurement time. The red spot corresponds to the area of the branching ratio $b$ and isomer transition energy $E_{\mathrm{is}}$ according to \cite{Beck09} (continued).}
\label{ff:3}
\end{figure}

To check whether a given set of experimental data corresponds to a single line or to a doublet, one can apply the likelihood ratio test~\cite{Pearson33, Wilks38, Wald43}. The essence of this test and the method to estimate the significance level is described in Appendix~A. Here we define the significance level as the probability to identify incorrectly the single peak as a doublet or the doublet as a single peak using the likelihood ratio test in the situation when we have either a single peak or a doublet whose parameters are specified. Figures ~\ref{ff:2},~\ref{ff:3}  show level curves for the signal count rates $R_{29}$ which are necessary to attain a significance level of 1\,\% for various values of the instrumental resolution $\FWHM$ and the $R_{29}/r_{\mathrm{bg}}$ ratio, in the $(b, E_{\mathrm{is}})$ plane. It is interesting to note that the optimal energy resolution is attained when $b \simeq 0.25$ and not for equally strong components of the doublet. This is explained by the fact that such a branching ratio leads to a noticeable asymmetry of the peak which facilitates the identification of a second component, increasing $b$ leads to a reduction of the main peak at fixed total signal count rate $R_{29}$.

\subsection{Precision in the determination of $E_{\mathrm{is}}$: Monte-Carlo simulations}

The aim of the proposed spectroscopy study is not only to resolve the 29.18\,keV line in the $\gamma$-spectrum of $^{233}$U as a doublet but to determine the energy splitting with maximum precision. 

We study the standard deviation
\begin{equation}
\delta E_{\mathrm{is}} = \sqrt{\langle (\hat{E}_{is}-E_{\mathrm{is}})^2 \rangle}  \label{ee:4}
\end{equation}
of the isomer energy as a characteristic measure of precision (in the following we call $\delta E_{\mathrm{is}}$ the {\em uncertainty of $E_{\mathrm{is}}$}). Here and below, angular brackets denote expectation values, $E_{\mathrm{is}}$ and $\hat{E}_{is}$ denote ``true'' and measured values of the isomer transition energy respectively. For the sake of brevity, we suppose that the true values of the energy splitting $E_{\mathrm{is}}$ and of the branching ratio $b$ are equal to 7.8\,eV~\cite{Beck09} and 1/14~\cite{Helmer94} respectively\footnote{According to~\cite{Beck07}, $b=1/13$ has 8\% error, therefore the value $b=1/14$ can be considered as a consistent but slightly pessimistic (from the point of view of spectral resolution) estimation.}. As before, we assume a total measurement time $t=10^6$\,s. Scaling the results to other values of $E_{\mathrm{is}}$, $b$, or $t$ is straightforward.

\begin{figure}
\begin{center}
\resizebox{0.99\textwidth}{!}{\includegraphics{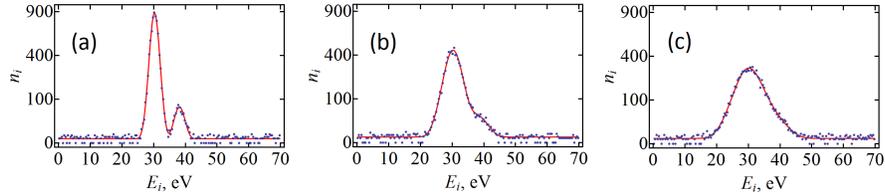}}
\end{center}
\caption{Examples of Monte-Carlo simulated ``experimental data'' (black dots) and fit (red curves) for $\FWHM=3$ eV (a), 6 eV (b) and 9 eV (c). Other parameters are: $R_{29}=7.74$ mHz, $r_{\mathrm{bg}}=3.9$ $\mu$Hz/eV, $t=10^{6}$ s. Plots are depicted in a ``square-root scale'' where the Poissonian noise is mapped onto signal-independent deviations. The origin of the energy axis is chosen arbitrarily.}
\label{fig:f1b}
\end{figure}

To investigate the dependence of $\delta E_{\mathrm{is}}$ on the instrumental energy resolution $\FWHM$, the signal count rate $R_{29}$, and the specific background count rate $r_{\mathrm{bg}}$, we perform a  Monte-Carlo study of $\delta E_{\mathrm{is}}$. For any set of parameters, we simulate the sample $\mathbf{n}$  as shown in Figure~\ref{fig:f1b}, and estimate the parameters $\theta=\{J_1, J_2, \tilde{\EE_1}, E_{\mathrm{is}}, \tilde{r}_{bg}, s\}$ maximizing the sum (\ref{ee:3}). We repeat this procedure  $10^4$ times and calculate $\delta E_{\mathrm{is}}$ according to (\ref{ee:4}).

\begin{figure}
\begin{center}
\resizebox{0.48\textwidth}{!}{\includegraphics{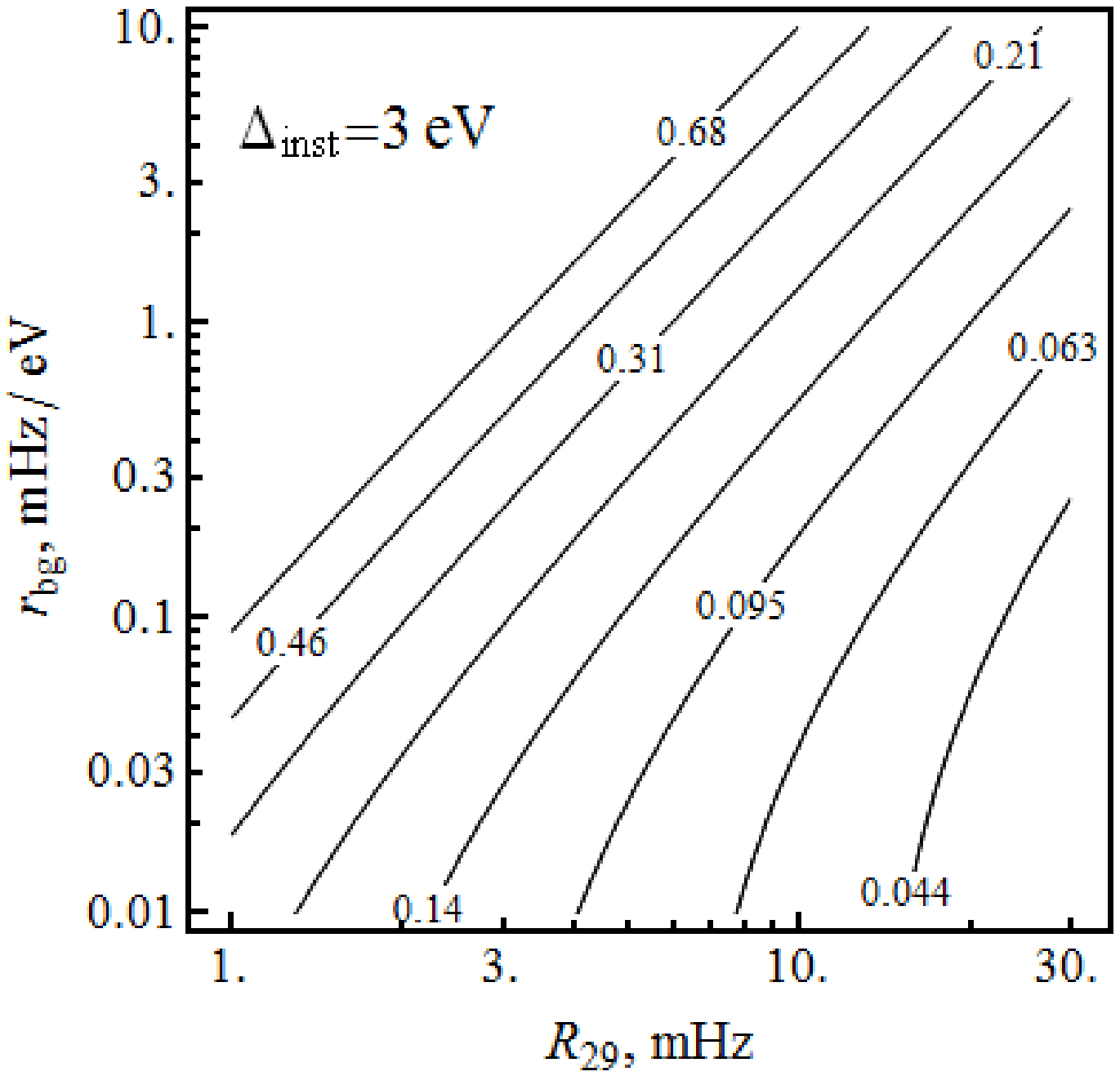}} \hspace{0.02\textwidth} 
\resizebox{0.48\textwidth}{!}{\includegraphics{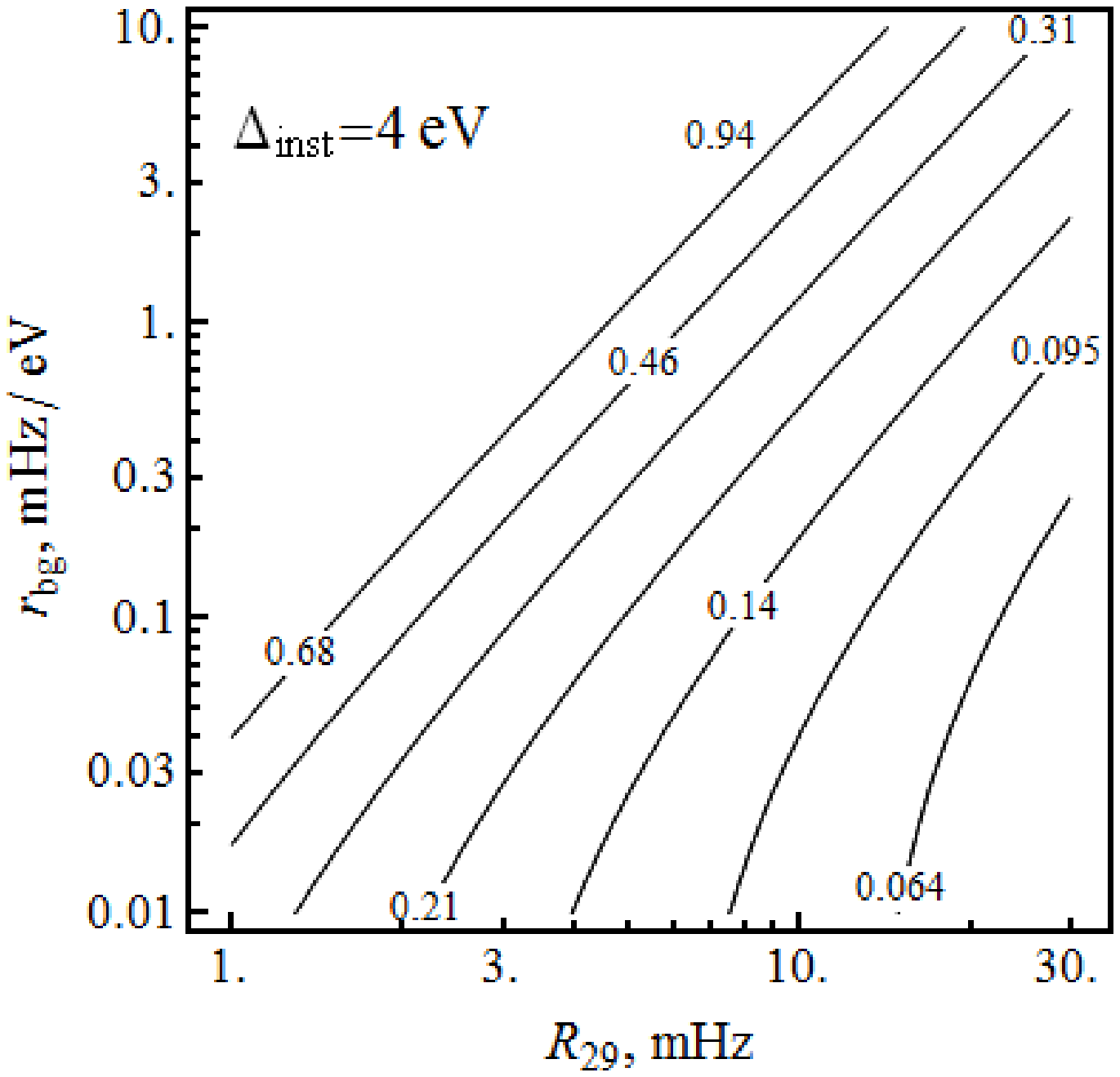}}\\
\resizebox{0.48\textwidth}{!}{\includegraphics{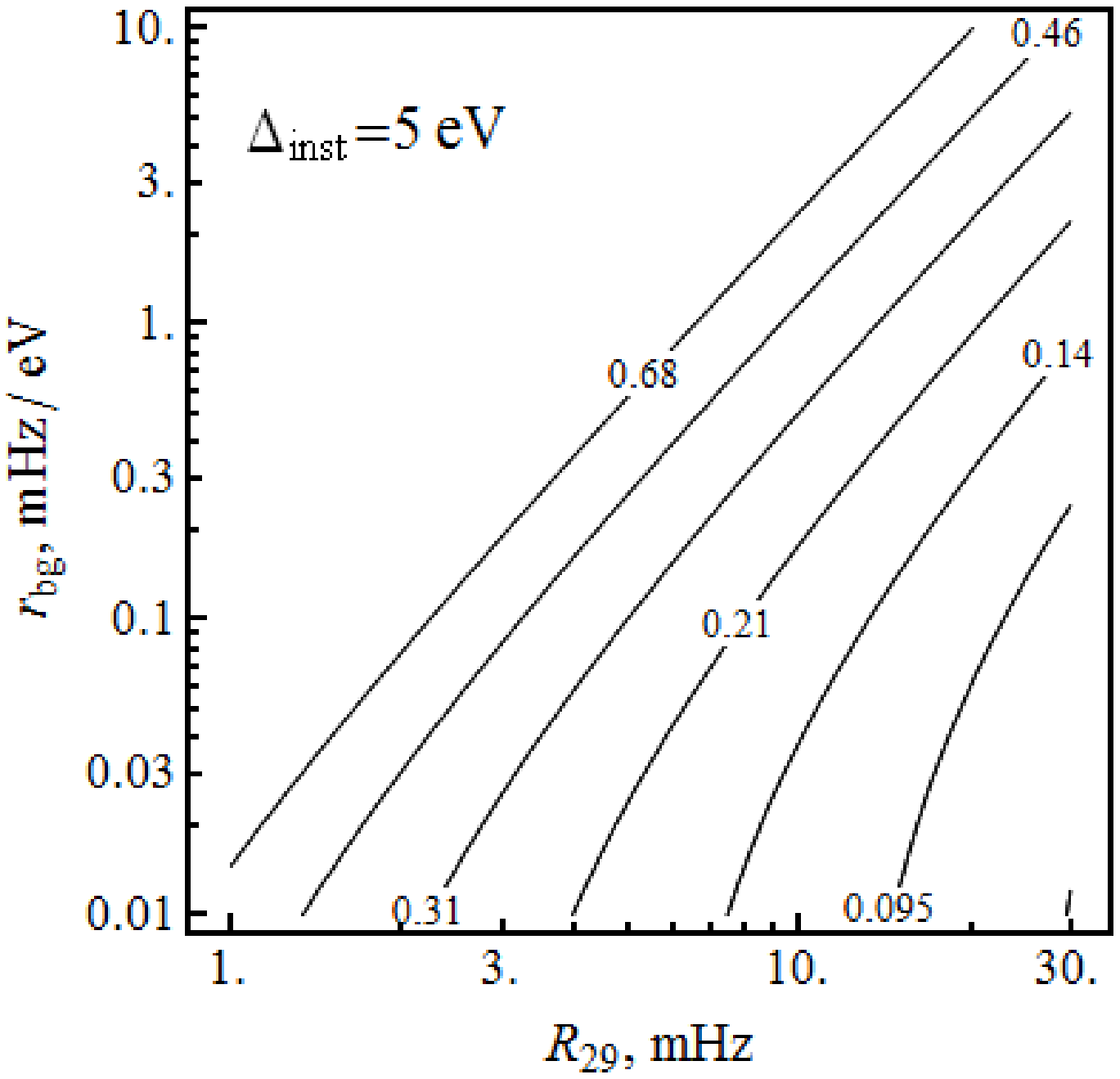}} \hspace{0.02\textwidth} 
\resizebox{0.48\textwidth}{!}{\includegraphics{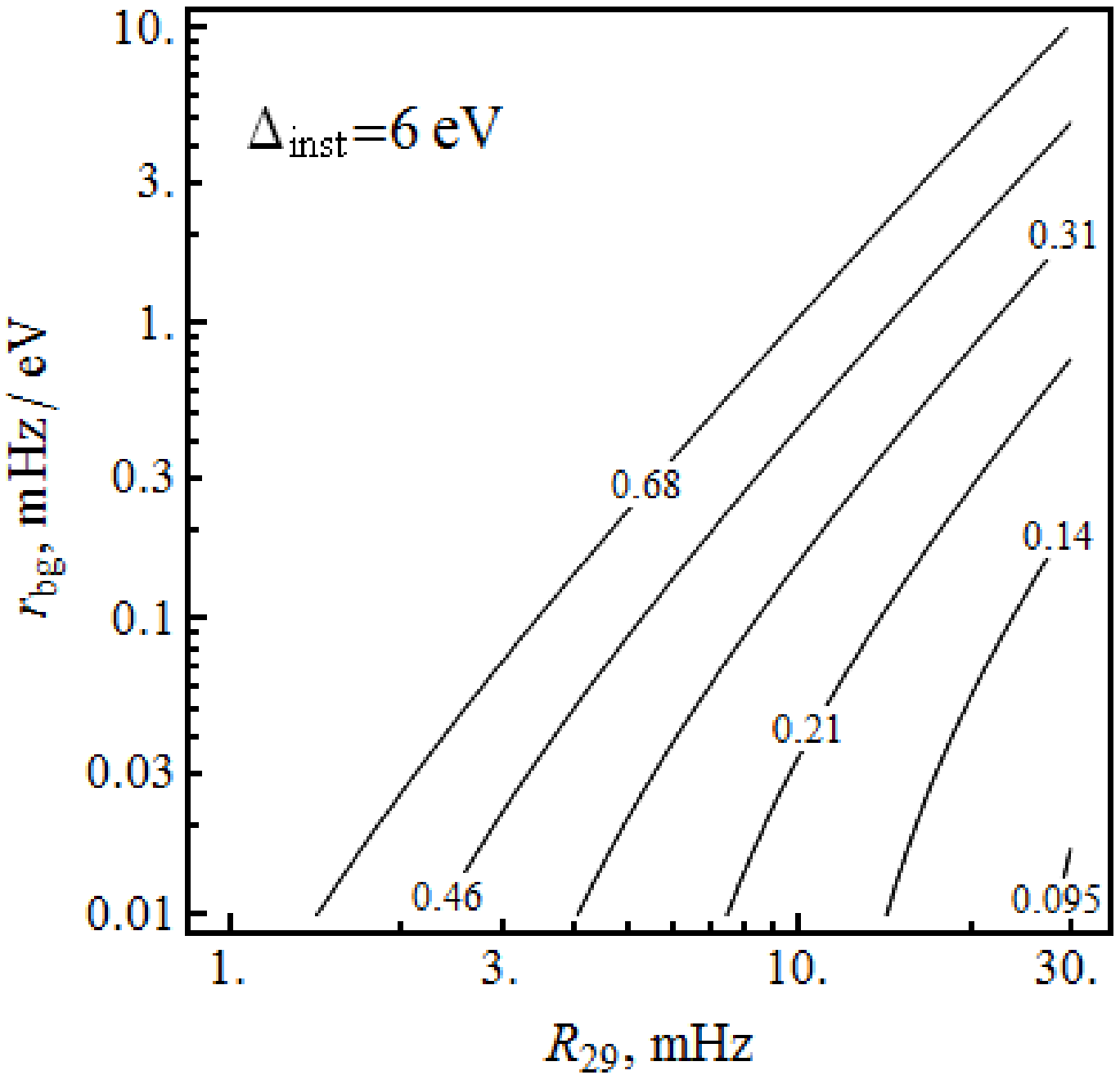}}\\
\resizebox{0.48\textwidth}{!}{\includegraphics{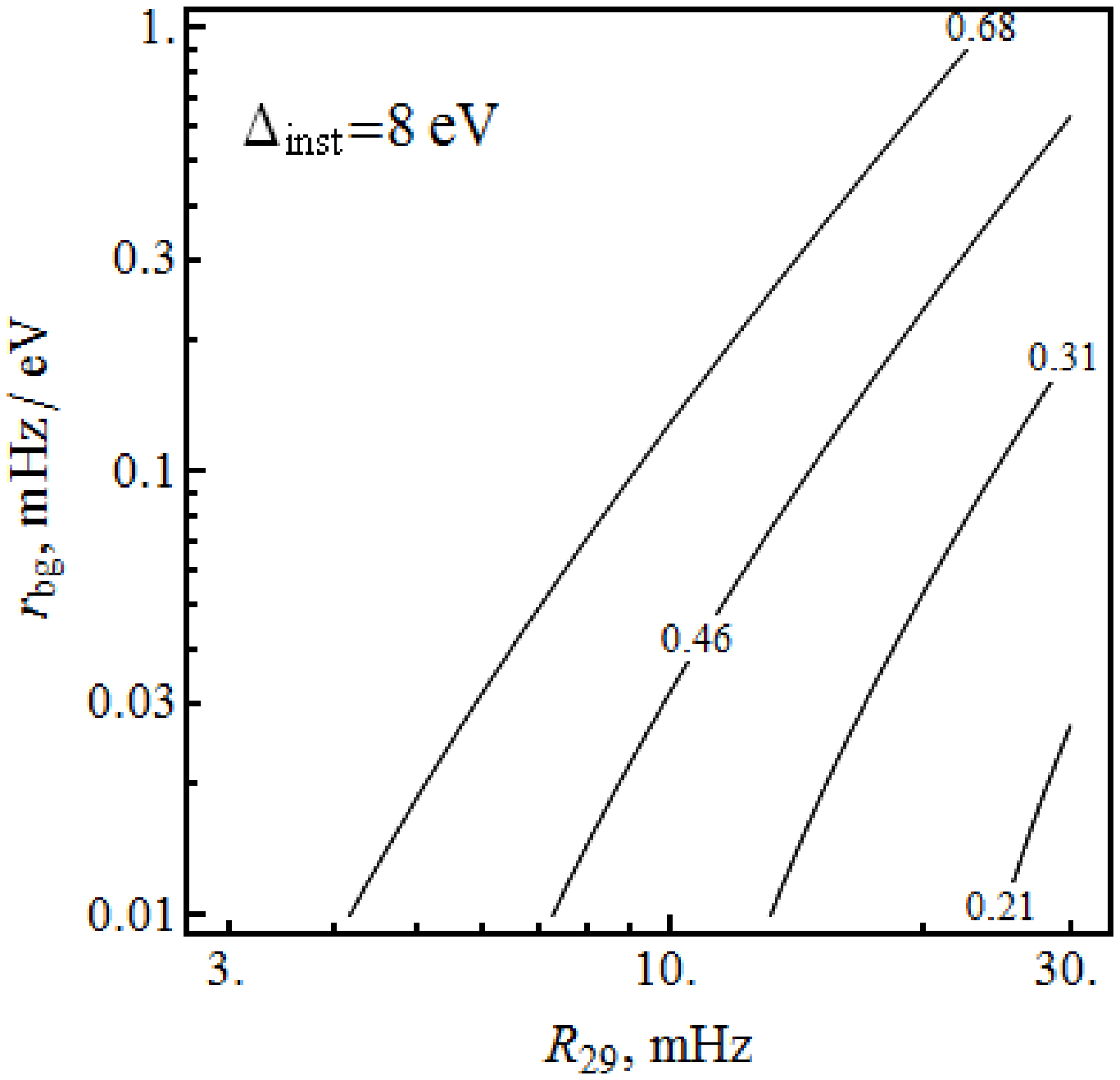}} \hspace{0.02\textwidth} 
\resizebox{0.48\textwidth}{!}{\includegraphics{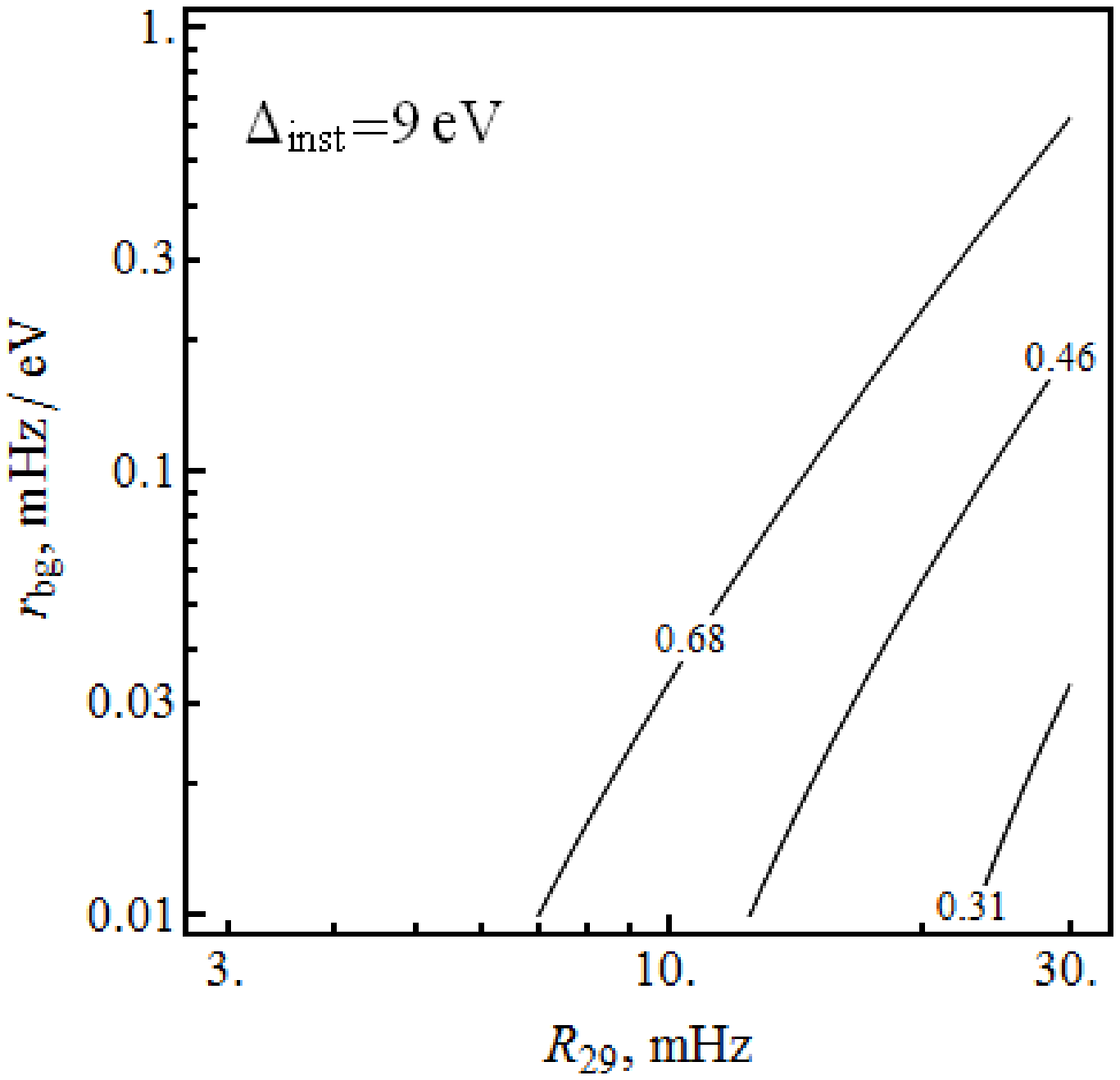}}
\end{center}
\caption{Curves of constant $\delta E_{\mathrm{is}}$ labeled in eV in the $(R_{29},r_{\mathrm{bg}})$ plane for different values of $\FWHM$ and $10^6$ s of total measurement time.}
\label{fig:f4}
\end{figure}

\begin{figure}
\begin{center}
\resizebox{0.48\textwidth}{!}{\includegraphics{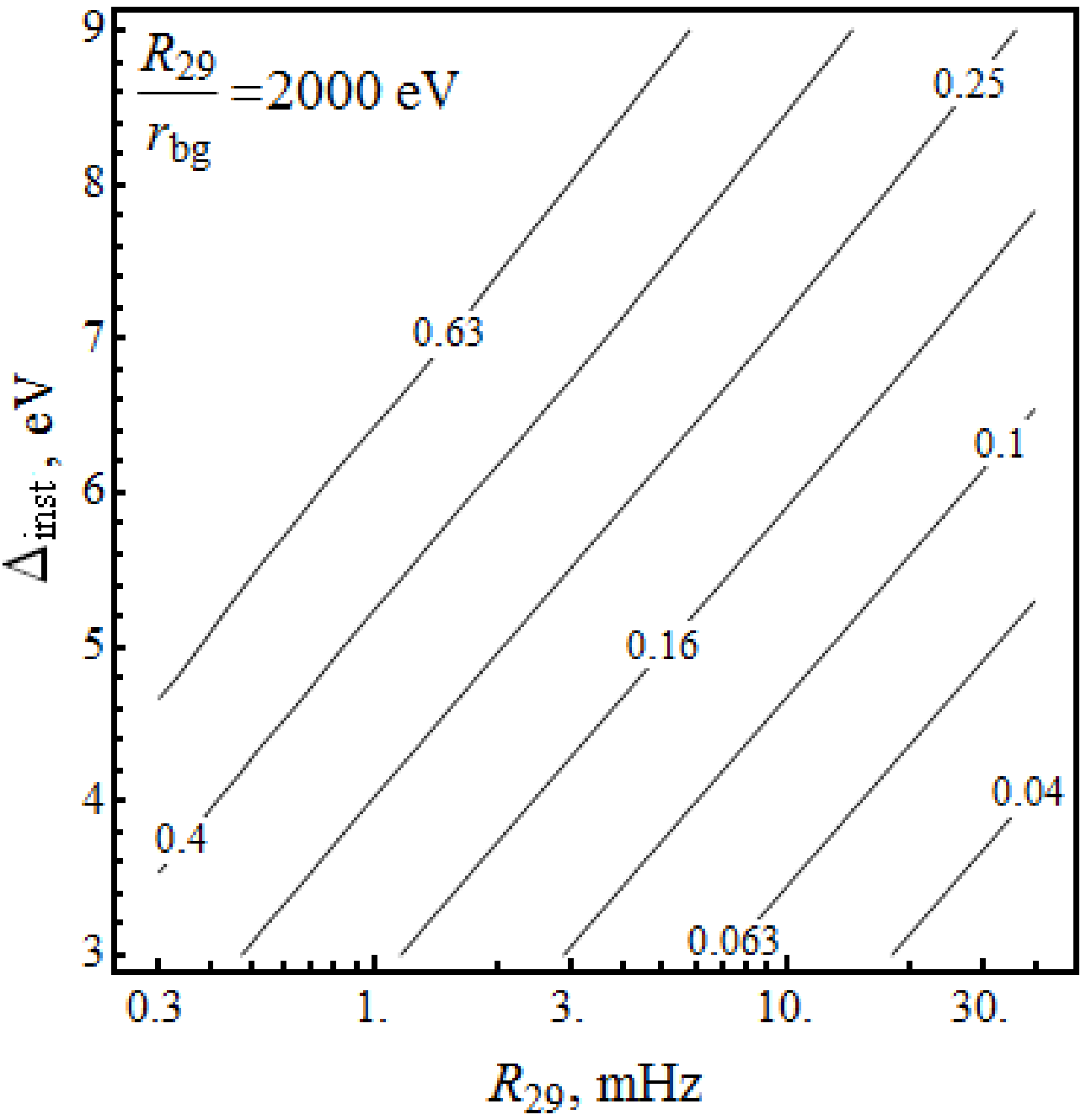}}
\resizebox{0.48\textwidth}{!}{\includegraphics{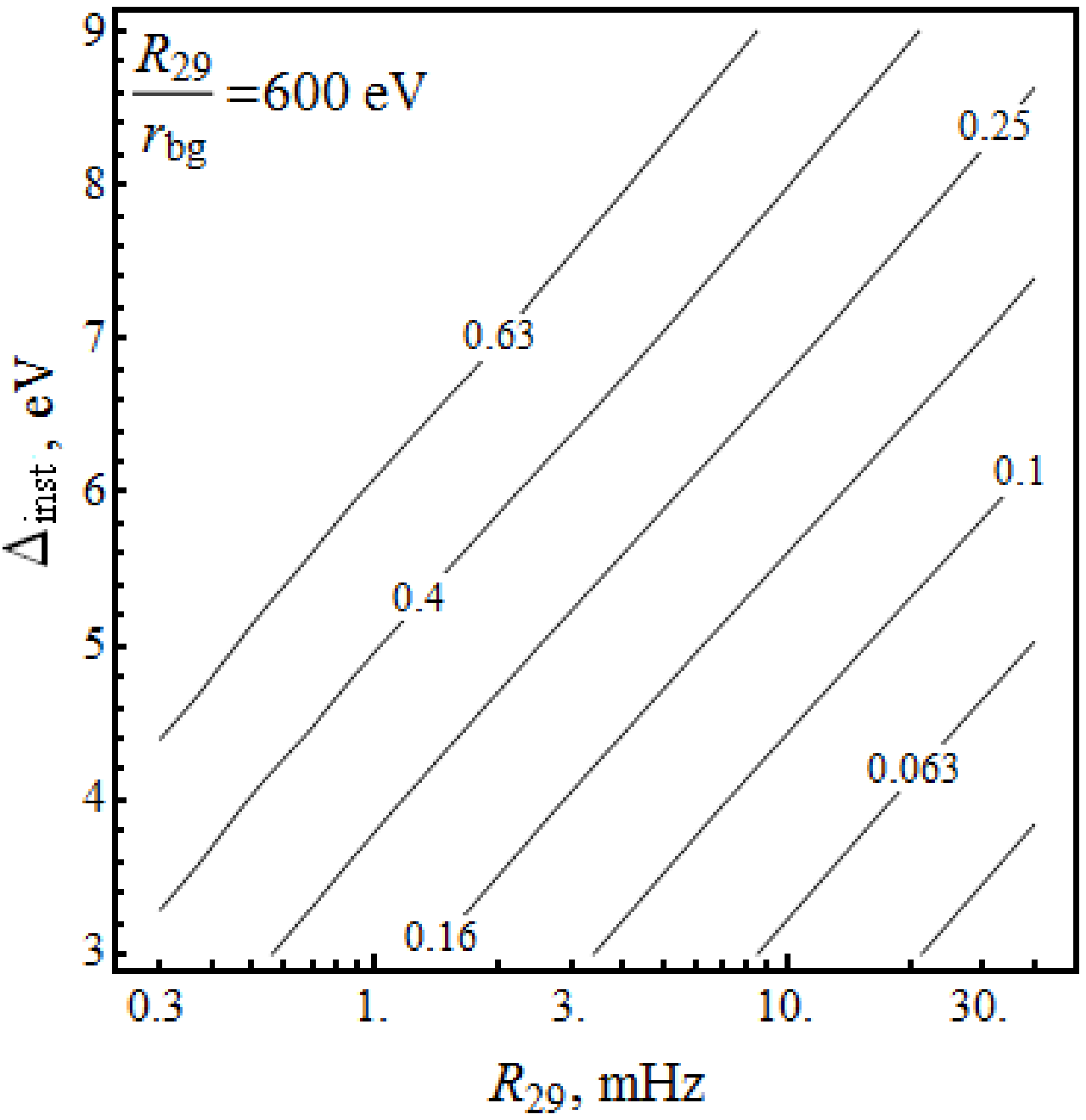}}
\resizebox{0.48\textwidth}{!}{\includegraphics{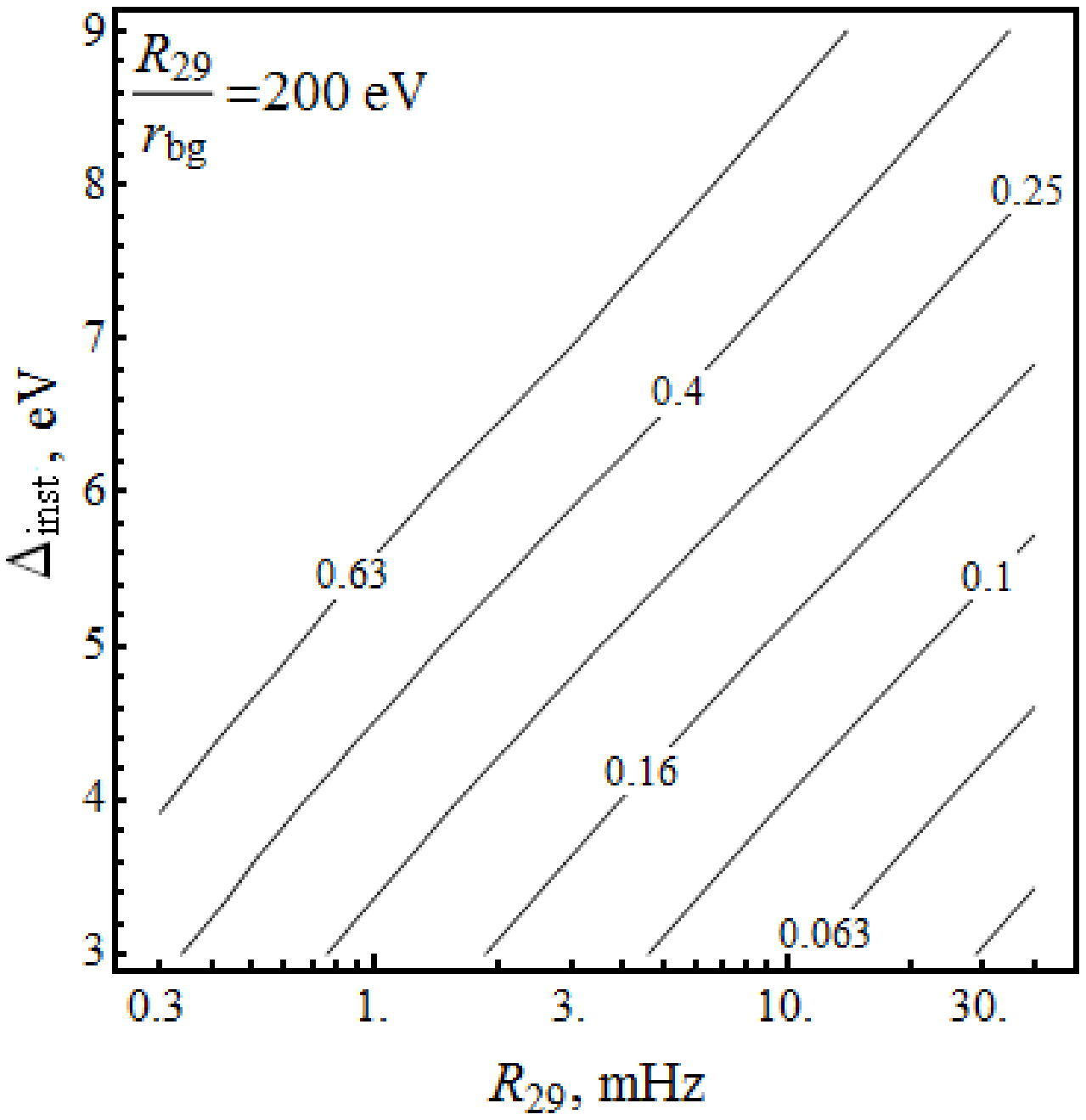}}
\resizebox{0.48\textwidth}{!}{\includegraphics{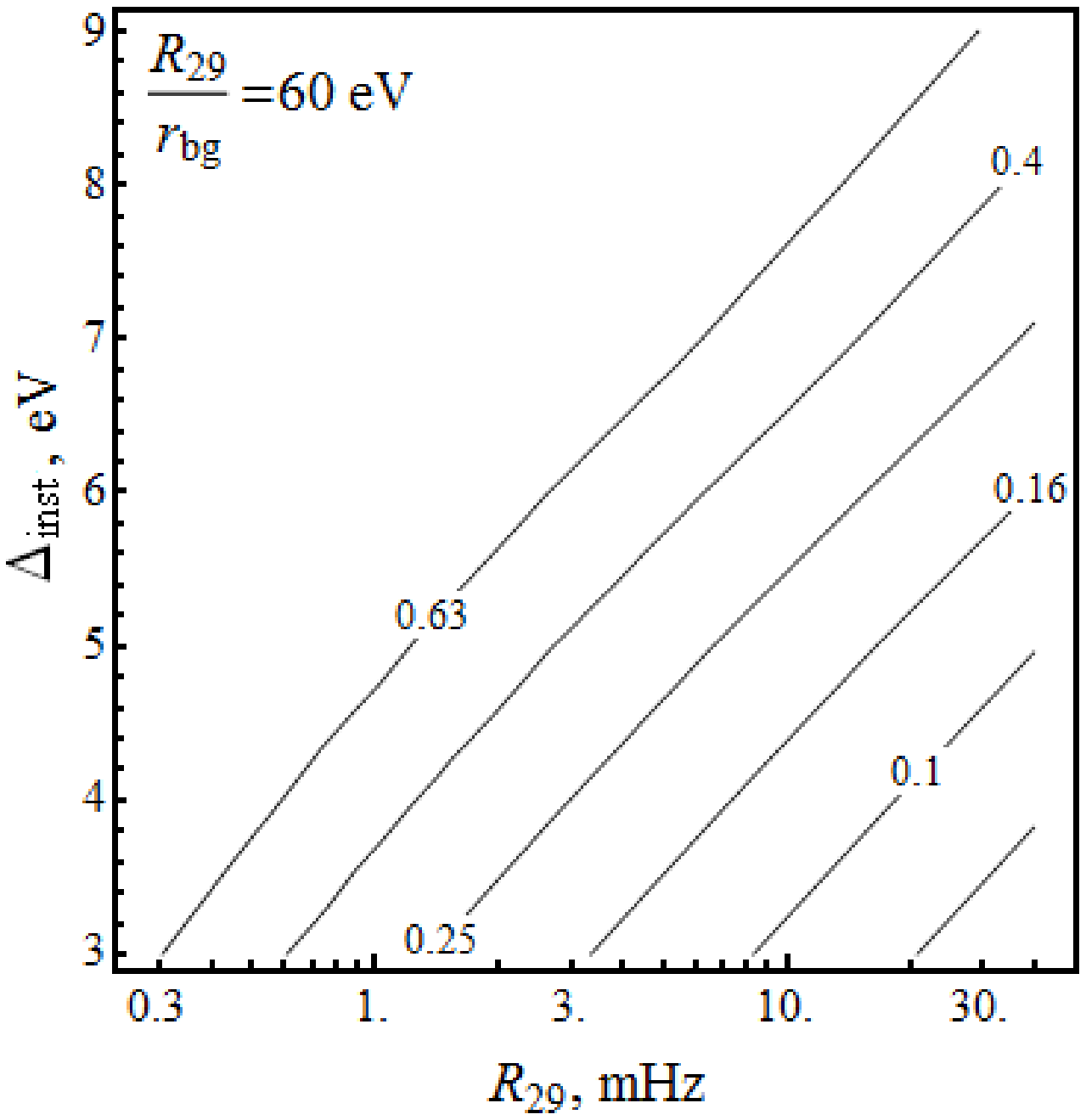}}
\end{center}
\caption{Curves of constant $\delta E_{\mathrm{is}}$  labeled in eV in the $(R_{29}, \FWHM)$ plane for different values of $R_{29}/r_{\mathrm{bg}}$ and $10^6$ s of total measurement time.}
\label{fig:f5}
\end{figure}

In Figure~\ref{fig:f4} we present curves of constant level of $\delta E_{\mathrm{is}}$ in the plane ($R_{29} $, $r_{\mathrm{bg}}$) obtained from the Monte-Carlo simulation. Finally, in Figure~\ref{fig:f5} we present the curves of constant $\delta E_{\mathrm{is}}$ in the plane $(R_{29}, \FWHM)$ for fixed ratios $R_{29}/r_{\mathrm{bg}}$. One can see that improving the instrumental resolution by 1\,eV increases the precision on the determination of $\delta E_{\mathrm{is}}$ by the same amount as doubling the 29.18\,keV signal count rate, or doubling the measurement time $t$.

\newpage
\section{Experimentally attainable count rates and expected precision}

In this section we estimate the attainable count rates and the resulting precision in a measurement of the isomer energy $E_{\mathrm{is}}$ that can be achieved with a state-of-the-art high-resolution microcalorimeter. In particular we consider the metallic magnetic microcalorimeter maXs-20 as described in ref.~\cite{Pies12}. The aim is to demonstrate, that valuable results can be obtained with currently available technology. In section~\ref{sec:developments} we describe ongoing work towards a more refined, dedicated detector setup.

The maXs-20 microcalorimeter consists of 8 detector elements (pixels) each of which has an absorber plate for incoming radiation ($250 \times 250 \times 5$\,$\mu$m$^3$ Au plate) connected to a $160 \times 160 \times 1.3$\,$\mu$m$^3$ paramagnetic temperature sensor (Er-doped Au) through 24 gold stems (10\,$\mu$m diameter and 5\,$\mu$m height each). Each sensor is connected to a thermal bath, the system is installed in a cryostat operating at a temperature of about 30\,mK. Energy deposited into the absorber plate heats the paramagnetic sensor and causes a change of its magnetization in an external magnetic field. Measuring this change in magnetization using SQUIDS, it is possible to determine the amount of absorbed energy and hence the energy of the incoming x- or $\gamma$-rays. Note that after the detection of an x- or $\gamma$-ray, the individual pixel can only detect again after a certain relaxation time of the order of 100~ms. Therefore the total count rate $R_\mathrm{T}$ should not be too high.

As a sample, we consider 1\,mCi of $^{233}$U electrodeposited as a film onto a metal planchet with a radius of $R=10$\,mm. We assume the sample to be situated 40\,mm from the detector (outside the cryostat) with a total surface $s=0.5$\,mm$^2$. Also we suppose the presence of additional material related to the cryostat vacuum system (sealing, input window of the cryostat, some other intentionally positioned shielding etc.), which we refer to as {\em filters}.

According to the NuDat 2.6 database~\cite{NUDAT}, each single decay of $^{233}$U is accompanied (on average) by one $\alpha$ particle with an energy from 4.309 to 4.824\,MeV, 0.213 conversion electrons with energies from 2.3 to 600\,keV (97.7\,\% of the electrons have energies below 50\,keV), and 0.0544 photons most of which (0.052 per decay) are $L$-shell x-rays with a mean energy of 13\,keV. As no individual $L$ x-rays are listed in NuDat 2.6 (only average energy and total intensity), we have taken the lacking data from the X-Ray Data Booklet~\cite{XRay}. 

To estimate the detector count rates, we suppose that all $\alpha$ particles and electrons emitted from the Uranium sample are stopped by the sample itself or by the filter materials, therefore, we consider only x- and $\gamma$-rays. Also it is supposed that all secondary electrons and photons generated in the filters are absorbed in the material locally. This assumption is correct for relatively thick filters made from light materials like Aluminium. 

We take into account absorption of the photons within the sample itself, the filters, and the detector. The total count rate $R_{T}$ is:
\begin{equation}
\begin{split}
R_{T}=&\sum_{i=1}^{i_{\mathrm{max}}}\frac{A\cdot \Omega}{4\pi}\cdot I_i \frac{1-e^{-\ell_\mathrm{U} a_\mathrm{U}(E_i)}}{\ell_U a_\mathrm{U}(E_i)} \\
& \times e^{- \ell_\mathrm{f} a_\mathrm{f}(E_i)}\cdot \left(1-e^{-\ell_\mathrm{Au}a_\mathrm{Au}(E_i)}\right). \label{ee:5}
\end{split}
\end{equation}
Here the sum is taken over all photon energies $E_{i}$, $A$ is the activity of the Uranium sample, $I_i$ is the relative intensity (quantum output) of photons with the energy $E_i$ per single decay event, $\ell_\mathrm{Au}$ and $\ell_\mathrm{f}$ are the thicknessess of the gold absorber and filters respectively. Linear absorption coefficients $a_{\kappa}(E_i)$ ($\kappa=\mathrm{U, Au, f}$) were taken from the XCOM Photon Cross Sections Database~\cite{XCOM}.
The count rate $R_{29}$ of signal photons is:
\begin{equation}
\begin{split}
R_{29}&=\sum_{i=1,2}\frac{A\cdot \Omega}{4\pi}\, I_i \frac{1-e^{-\ell_\mathrm{U} a_\mathrm{U}(E_i)}}{\ell_\mathrm{U} a_\mathrm{U}(E_i)} \, e^{- \ell_\mathrm{f} a_\mathrm{f}(E_i)}\times \\
& \left[1-e^{-\ell_\mathrm{Au}a_\mathrm{Au}(E_i)} - \varpi \sum_X I_X P_\mathrm{es}(E_i,E_X,\ell_\mathrm{Au}) \right], \label{ee:6}
\end{split}
\end{equation}
where $i=1,2$ corresponds to the two components of the doublet, $\varpi=0.331$ is the Au $L$ shell fluorescence yield \cite{Hubbell94}, $I_X$ is the probability that an energy of a fluorescence photon emitted by a Au atom is equal to $E_X$, and $P_\mathrm{es}(E_{\gamma},E_X,\ell_\mathrm{Au})$ is a probability that an incoming $\gamma$-quant with energy $E_i$ will be absorbed, and an $x$-ray photon following this absorption leaves the absorber (escape line). Supposing an isotropic spatial distribution of these secondary photons, we obtain
\begin{equation}
\begin{split}
P_\mathrm{es}=&\frac{1}{2} \int\limits_{0}^{\ell}a_\gamma  e^{-a_\gamma  x} 
\left(\int\limits_{0}^{\pi/2} \exp\left[ \frac{-a_X(\ell-x)}{\cos\theta}\right]\sin \theta d\theta\right. \\
&\left.+\int\limits_{\pi/2}^{\pi} \exp\left[ \frac{a_X x}{\cos\theta}\right] \sin \theta d\theta \right) dx, \label{ee:7}
\end{split}
\end{equation}
where $\ell=\ell_\mathrm{Au}$, $a_\gamma=a_\mathrm{Au}(E_\gamma)$, $a_X=a_\mathrm{Au}(E_X)$. Also we suppose that the $x$-ray photon is emitted from the $L$ shell, i.e. the deepest shell that is accessible by energy conservation, and the probability $I_X$ for emission of the photon is the relative intensity  tabulated in~\cite{XCOM} normalized to the sum of relative intensities from the $L$ shell.

Evaluating the expressions (\ref{ee:5}) -- (\ref{ee:7}) yields a total detector count rate $R_\mathrm{T}=1$\,Hz and a signal count rate of $R_{29}=7.74$\,mHz for a 1.3\,mm thick Aluminium filter. Without any filter, the count rates for the same parameters are: $R_\mathrm{T}=13.6$\,Hz, and $R_{29}=11.78$\,mHz. We see that the Aluminium filter absorbs approximately 92\,\% of all photons emitted from the sample, but only about 34\,\% of the signal photons. We conclude that filtering is an effective method to decrease the total count rate $R_\mathrm{T}$, caused mainly by low-energy Thorium $L$ shell x-ray.

The background count rate $r_{\mathrm{bg}}$ is caused by the escape of some fraction of the dissipated energy of $\gamma$-quants absorbed in the detector. In~\cite{Beck07}, the number of background counts close to the 29.18\,keV doublet was about 30 -- 40 events per 3\,eV bin whereas the total number of counts in the 29.18\,keV peak was about $2.7 \cdot 10^4$ events. This yields the ratio $R_{29}/r_{\mathrm{bg}}=2000$\,eV. Assuming that a similar ratio can be realized with the maXs-20 detector, we find that the uncertainty $\delta E_{\mathrm{is}}$ on the measured isomer transition energy $E_{\mathrm{is}}$ will be equal to 0.06\,eV for an instrumental resolution of $\FWHM=3$\,eV, signal count rate $R_{29}=7.74$\,mHz and total measurement time $t=10^6$ s. Therefore the proposed experiment to determine the isomer energy $E_{\mathrm{is}}$ is expected to be almost one order of magnitude more precise than the results obtained in the previous experiment~\cite{Beck07}. Reducing the experimental resolution to 6\,eV and 9\,eV yields $\delta E_{\mathrm{is}}=0.19$\,eV and $\delta E_{\mathrm{is}}=0.56$\,eV respectively. Increasing the total measurement time to $t=2.6\cdot 10^6$\,s, 1 month, we can measure the isomer transition energy with an uncertainty $\delta E_{\mathrm{is}}=0.037$\,eV for $\FWHM=3$\,eV, $\delta E_{\mathrm{is}}=0.12$\,eV for $\FWHM=6$\,eV, or $\delta E_{\mathrm{is}}=0.33$\,eV for $\FWHM=9$\,eV.

\section{Further statistical aspects}
\label{sec:issues}
We are aware of certain simplifications and assumptions in the above analysis. Here we briefly resume some additional issues that could arrise, a detailed discussion of these points is beyond the scope of this work.

First, the shape of signal peaks can deviate from Gaussian. For example, a long low energy tail on the spectral lines may lead to the appearance of a noticeable step in the backgound count rate (see, for example, Figure~2 (a) in ~\cite{Beck07}). We believe that in the work of ref.~\cite{Beck07} this effect is caused mainly by the escape of energy from the absorber material, for example in the form of athermal phonons~\cite{Pies12}. The yield of Compton scattering is not sufficient to explain this step, see Appendix~B. To take this effect into account correctly, we will have to modify our model of the background. A more difficult situation arises when the escaping energy is relatively small, which would lead to an asymmetry of the line rather than the appearance of a tail. In this case, it would be useful to study an isolated single peak separated from the doublet of interest but intense enough to give good statistics, and/or to perform an independent study with another $\gamma$ source, e.g. $^{241}$Am.

Another issue that may appear is a slow time-dependent fluctuation of the response function caused by an uncontrollable drift of ambient magnetic fields and/or cryostat temperature over the duration of the measurement. We believe we can suppress such drifts below 10\,eV by temperature stabilization and mu-metal shielding of the setup. Additionally, we will monitor the position of a series of reference x- and $\gamma$-ray lines for a correct tracing of this drift, realizing a time-dependent calibration of the detector.  Note that x-ray lines generally have a much broader linewidth than $\gamma$ lines~\cite{Raboud99} which simplifies the identification. An auxiliary calibration source, for example $^{241}$Am, can help to enhance the quality of this callibration.

Also we should mention possible interference of the 29.18\,keV doublet with coincidence and escape lines of x- and $\gamma$-rays of $^{223}$U and other elements present in the sample. We plan to study the composition of the sample using ``ordinary'' low-precision $\gamma$-spectrometry.

\section{Planned experimental implementation}
\label{sec:developments}

\subsection{Detector development}
We are currently developing a dedicated new detector for the measurement described above, to some extend interpolating between the maXs-20 (0-20\,keV) and the maXs-200 (0-200\,keV) series~\cite{Pies12}. It will feature a linear array of magnetic calorimeters, each with an active area of $250 \times 250$\,$\mu$m$^2$. We will increase the absorber thickness by a factor of 2 to 3 in comparison with 5\,$\mu$m in the maXs-20, leading to a stopping power of about 50\,\% at 30\,keV. We will operate the detectors in a dry 3He/4He-dilution refrigerator at about 20~mK. In this situation, the intrinsic energy resolution of the detector caused by thermal noises of all kinds is expected to be below 2\,eV (FWHM). According to the calorimetric detection principle of metallic magnetic calorimeters~\cite{Fleischmann05}, this resolution is independent of energy as long as the total gain (including operational temperature and external magnetic fields) is stable, the dependence of the detector response on the event position in the absorber is negligible, and the photon energy $E$ is still small enough to be within the range of linear detector response $\Phi(E)$. Also we expect that the minimal time between two correctly measurable counts in a single detector element (pixel) will be about 100 ms.

\begin{figure}[t]
\begin{center}
\resizebox{0.8\textwidth}{!}{\includegraphics{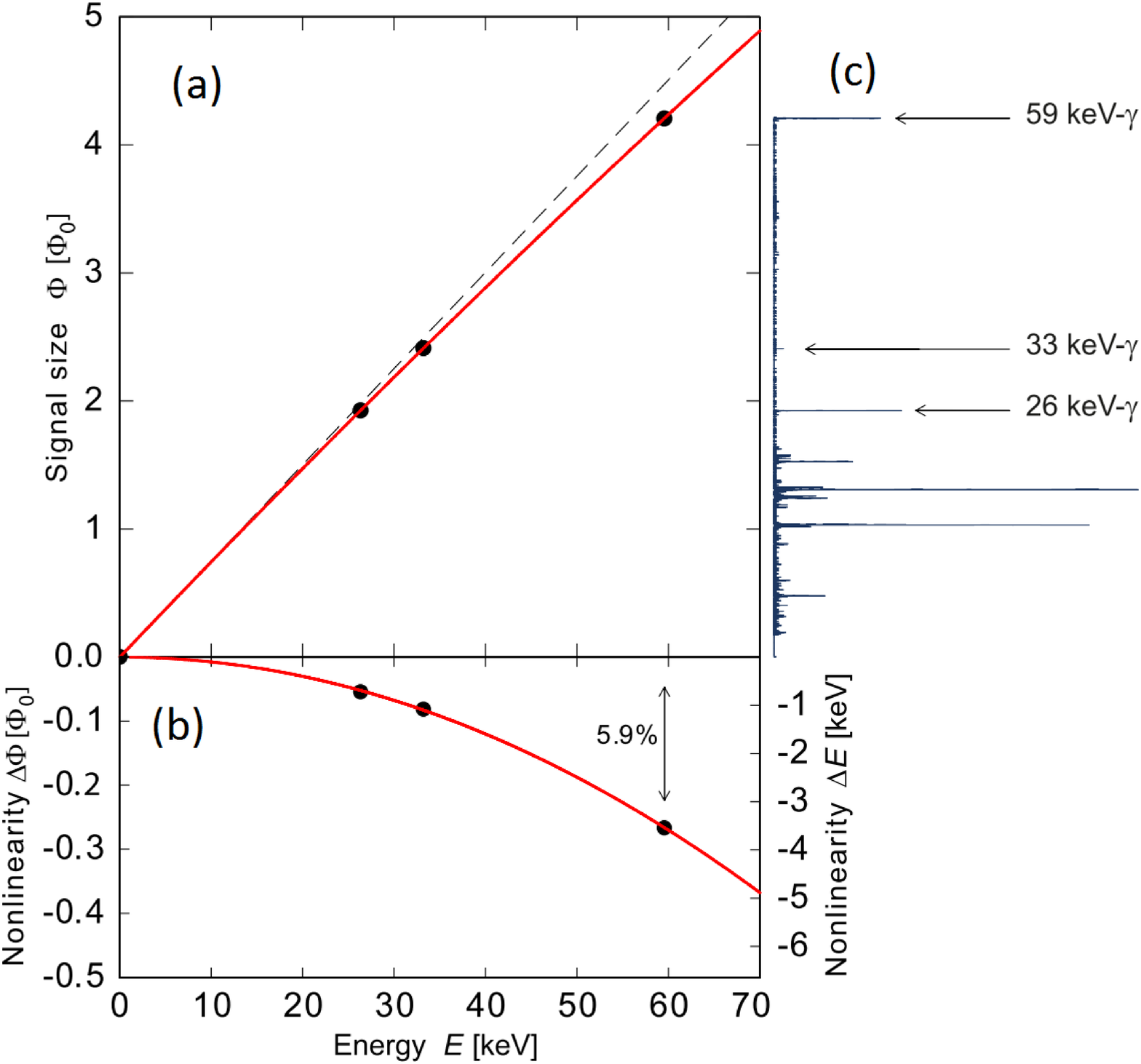}}
\end{center}
\caption{Detector response versus the energy of absorbed photons (a) and deviation from the linear behaviour (b) based on the measurement of 3 characteristic lines of an $^{241}$Am $\gamma$-spectrum (c).}
\label{fig:7}
\end{figure}

So far, we have achieved resolving powers up to about $E/\FWHM =3700$ (corresponding to 1.6\,eV (FWHM) at 5.9\,keV) with our maXs-20 devices (unpublished), being limited by a combination of both, instabilities of the operating temperature and a position dependence. We believe that we can improve the short-term stability of the total gain and keep the position dependence of absorbtion events small enough to allow resolving powers beyond $10^4$ in the planned experiment. Also, the response of the present maXs-20 detector to photon energies below 60\,keV has a small quadratic deviation from a linear behavior, see Figure~\ref{fig:7}. At an energy of 30\,keV this deviation is only about 3\,\%, which yields a 6\,\% degradation of the intrinsic energy resolution compared to the low-energy signals, i.e. below 2.12\,eV on an absolute scale.

\subsection{Sample preparation and characterization}

\begin{figure}[t]
\begin{center}
\resizebox{0.98\textwidth}{!}{\includegraphics{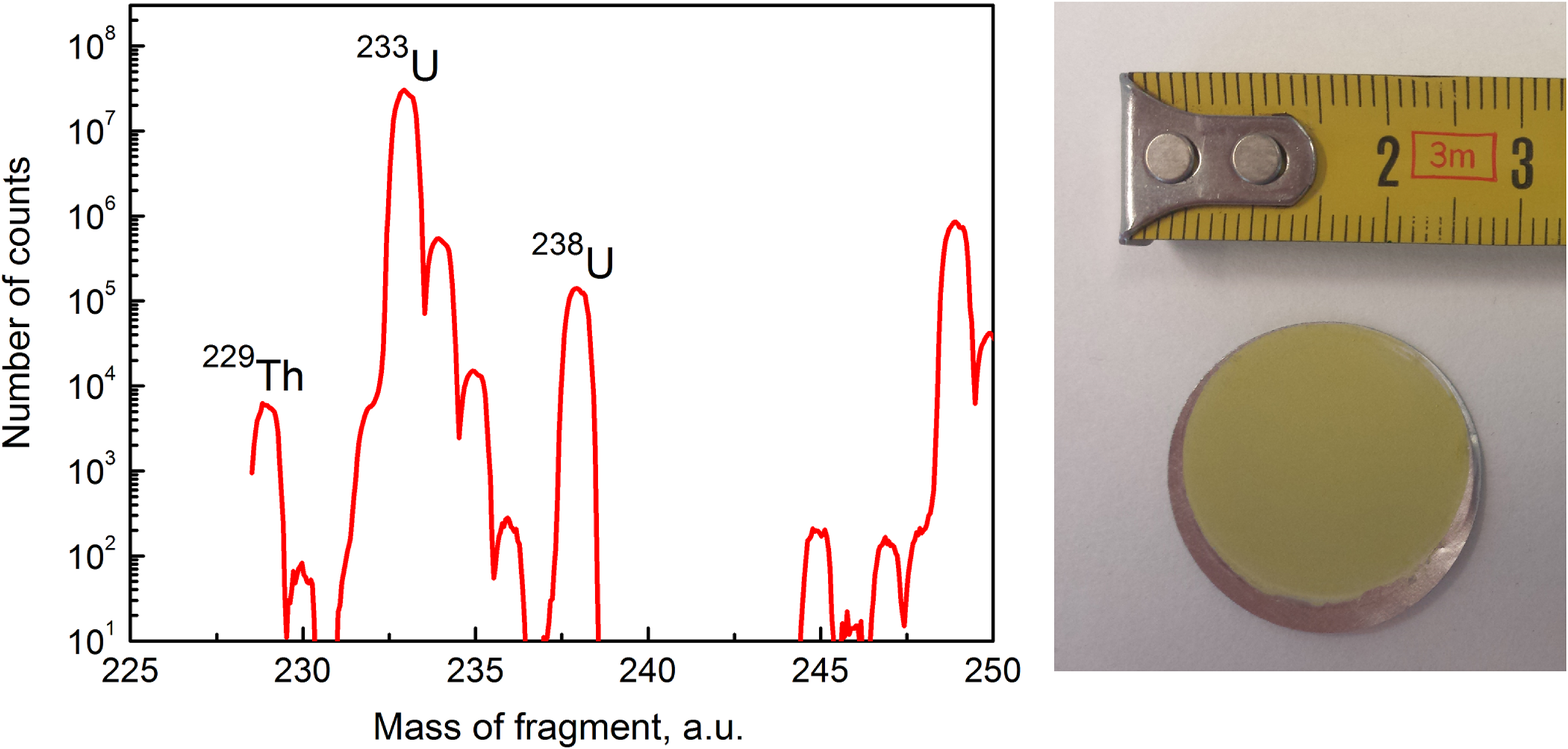}}
\end{center}
\caption{Left: Inductively coupled plasma mass spectrometer (ICP-MS) data of the raw $^{233}$U material composition. Mass signals above 240 amu are molecular fragments and can be ignored. Right: Photo of the electroplated UO$_2$ test sample (with $^{238}$U).}
\label{fig:8}
\end{figure}

The sample should ideally consist of isotopically pure $^{233}$U to avoid a too high count rate not carrying relevant information and possible interference with the 29.18\,keV Thorium doublet signal. For this project, we have 560 mg (about 5\,mCi activity) raw sample material available (in oxide powder form). The origin and preparation procedure of this material is unknown, from the $\gamma$-spectrum we suspect that originally, $^{232}$Th has been activated in a high-flux neutron reactor and the $^{233}$U has been separated chemically. A mass spectrum of the raw material, produced by an in-house ICP-MS can be seen in Figure~\ref{fig:8}. The raw material contains $> 90$\,\% $^{233}$U, together with traces of $^{232}$U, $^{234}$U, $^{235}$U, $^{238}$U, and the decay product $^{229}$Th. Further daughter product of the $^{233}$U chain have not been detected. 

To further purify the sample, we will perform a PUREX Uranium extraction procedure. We have also observed an efficient additional element separation in the electrodeposition process. In-house analysis using $\gamma$- and $\alpha$-spectroscopy, ICP-MS, and neutron activation analysis will allow us to quantify the success of this procedures and finally know the exact composition of the final measurement sample. 

The sample will be produced by electroplating $^{233}$U from a liquid solution onto a stainless steel or aluminium planchets. The target activity of 1\,mCi corresponds to 104\,mg of pure $^{233}$U or 118\,mg of UO$_2$. Producing correspondingly thick films (15-20\,$\mu$m) turned out to be difficult in electroplating \cite{Hallstadius84, Luna99}. We have therefore developed a process to deposit up to 20\,mg Uranium onto stainless steel or aluminium foils of only 10-50\,$\mu$m thickness. These samples can easily be stacked to realize the target activity without the carrier foils significantly reducing the count rate in the 29.18\,keV peak.

\section{Conclusion}
We have analyzed the feasability of an indirect measurement of the low-energy isomer state in $^{229}$Th using a high-resolution magnetic microcalorimeter. We propose to resolve the 29.18\,keV doublet in the $\gamma$-radiation spectrum following the $\alpha$-decay of $^{233}$U. Such a measurement would provide a strong indication for the existence of the isomer state and improve the accuracy on the energy measurement significantly. The measurement appears feasable with currently available detector technology and samples.

\section{Acknowledgements}
We thank C.~Streli, P. ~Wobrauschek, and J. ~Seres for discussions and calculations concerning $\gamma$-ray filters. We thank S.~Smolek and M.~Gollowitzer for preliminary tests and characterizations of $^{233}$U samples. This work was supported by the ERC project 258604-NAC, the FWF Project Y481, the WPI Thematic Program ``Tailored Quantum Materials'', and the FWF project M1272-N16 ``TheoNAC''.

\newpage
\section*{Appendix A. Likelihood ratio test and estimation of the significance level $a$}
Here we describe the essence of likelihood ratio tests for regression models in the simple case of normally distributed observables with known dispersions, and the method which we actually used for the estimation of the significance level attainable in the experiment.

Let us have $N$ experimental observables $y_i$ normally distributed around their (unknown) expectation values $\lambda_i$ with known dispersions $\sigma_i$. Without lost of generality, we can set $\sigma_i=1$ for all $i$. Also, we have 2 regression models, one of which ({\em short model}) is a particular case of another one ({\em long model}). In the long model, it is supposed that the expectation values of observations $y_i$ are some known functions $f_i(\theta)$ of the $l$-dimensional parameter $\theta=\{\theta_1,...,\theta_l\}$. In the short model, it is supposed also that the parameters $\theta$ has some additional restrictions, and the short model has $s=l-r$ degrees of freedom. It is supposed that the long hypothesis anyhow is correct. The question we want to answer is whether the short hypothesis is correct? Or, more precisely: how plausible (or unplausible) is it to obtain the set of observables $y_i$, if the short model is correct?

\begin{figure}
\begin{center}
\resizebox{0.48\textwidth}{!}{\includegraphics{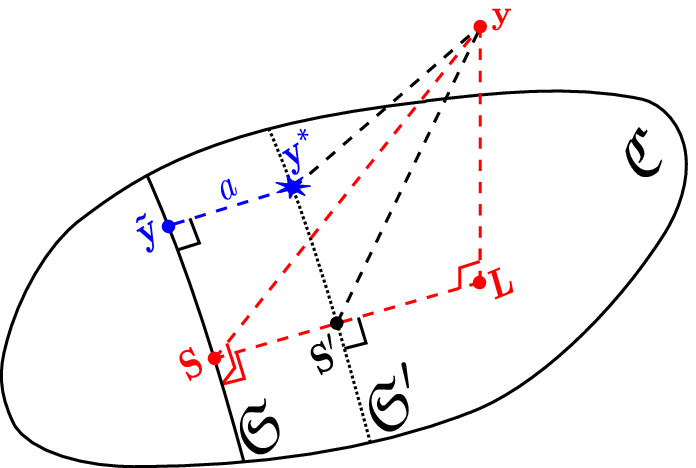}}
\end{center}
\caption{Sketch of the sample space: $\CC$ and $\SS$ are the long- and short hypotheses surfaces respectively. Point $\mathbf{y^*}=\mathbf{f}(\theta^*)$ corresponds to the true value of parameters, $\SS^\prime$ is an $s$-dimensional surface passing through $\mathbf{y^*}$ whose points are equidistant to $\SS$. $\mathbf{y}$ is an experimental point, $\mathbf{L}=\mathbf{f}(\theta^L)$, and $\mathbf{S}=\mathbf{f}(\theta^S)$ corresponds the best fits within the long and short hypotheses respectively. The non-centrality parameter is the square of the distance $a$ between $\SS$ and $\SS^\prime$.}
\label{fig:A1}
\end{figure}

The likelihood ratio test is a powerful method to answer this question. To illustrate the essence of this test, let us represent $N$ observations $y_i$ as a point $\mathbf{y}$ in the $N$-dimensional Euclidean sample space. $N$ functions $f_i(\theta_1,...,\theta_l)$ form an $\l$-dimensional surface $\CC$ corresponding the long hypothesis. In turn, these functions with additional conditions corresponding the short hypothesis form the $s$-dimensional surface $\SS$ within $\CC$, see Figure~\ref{fig:A1}. We suppose that these surfaces are sufficiently smooth. In the case of linear regression models, the functions $f_i(\theta)$ are linear, and the surfaces $\CC$ and $\SS$ are just hyperplanes. 

For normally distributed observables $y_i$ with zero mean and unit dispersion and for some set of parametes $\theta$, the logarithmic likelihood function is 
\begin{equation}
\ell(\mathbf{y}|\theta)=const-\frac{1}{2}\sum_{i=1}^N (f_{i}(\theta)-y_i)^2.
\label{ee:a1}
\end{equation}
Therefore the maximization of $\ell(\theta)$ is equivalent to a minimization of the distance between the points $\mathbf{f}(\theta)$ and $\mathbf{y}$ in the sample space. The square of this distance we denote as $|\mathbf{f}_{\theta}-\mathbf{y}|^2$. Let $\mathbf{L}=\mathbf{f}(\theta^c)$ and $\mathbf{S}=\mathbf{f}(\theta^s)$, where $\theta^c$ and $\theta^s$ are the best likelihood estimations of parameters $\theta$ within the long and short hypotheses respectively. It is easy to see that $|\mathbf{y}-\mathbf{S}|^2\simeq |\mathbf{y}-\mathbf{L}|^2+|\mathbf{L}-\mathbf{S}|^2$ which yields $|\mathbf{L}-\mathbf{S}|^2=2 \big(\ell(\mathbf{y}|\theta^c)-\ell(\mathbf{y}|\theta^s)\big)$. In turn, $|\mathbf{L}-\mathbf{S}|^2$ is a sum of squares of $r$ normally distributed random values $\kappa_\alpha$ with unit dispersion and different means $\mu_\alpha$. Therefore, $|\mathbf{L}-\mathbf{S}|^2$ is a non-central $\chi^2$ random value with $r$ degrees of freedom and non-centrality parameter $a^2=\sum_{\alpha=1}^r \mu_\alpha^2$. For the sake of brevity, we denote this random value as $\chi^2_r(a^2)$. It is easy to see that $a$ is just a distance between the short hypothesis surface $\SS$ and the point $\mathbf{y^*}=\mathbf{f}(\theta^*)$ corresponing to the true value $\theta^*$ of parameters $\theta$, see Figure \ref{fig:A1}. If the short hypothesis is true (it corresponds the situation when the surface $\SS$ coincided with $\SS^\prime$ in Figure \ref{fig:A1}), then $a=0$ and $|\mathbf{L}-\mathbf{S}|^2$ is just a ``usual'' $\chi^2_r$ random value with $r$ degrees of freedom. 

To test whether the short hypothesis is true or not, one should choose some desirable significance level $\alpha$, and compare the value
\begin{equation}
D=|\mathbf{L}-\mathbf{S}|^2=2(\ell(\mathbf{n}|\,\theta^c)-\ell(\mathbf{n}|\,\theta^s))
\label{ee:a2}
\end{equation}
with some critical value $\lambda_\alpha(\chi^2_r)$ such, that the probability 
\begin{equation}
P\big(\chi^2_r>\lambda_\alpha(\chi^2_r)\big)=\alpha.
\label{ee:a3}
\end{equation}
If $D>\lambda_\alpha(\chi^2_r)$, the short hypothesis is rejected on significance level $\alpha$, otherwise it is accepted. The probability to reject the short hypothesis incorrectly is equal to $\alpha$. On the other hand, the probability to accept the short hypothesis incorrectly is $P(\chi^2_r(a^2)<\lambda_\alpha(\chi^2_r))$. Therefore if the non-centrality parameter $a^2$ will be larger than $a_r^2(\alpha)$ such that 
\begin{equation}
P\big[\chi^2_r(a_r^2(\alpha))<\lambda_\alpha(\chi^2_r)\big]=\alpha,
\label{ee:a4}
\end{equation}
the probability to accept the short hypothesis falsely will be less than $\alpha$.

In our case, the observables $n_i$ are not normal random values with known dispersion but Poissonian random values. To estimate the possibility to identify the 29.18\,keV line as a doublet, we consider$\sqrt{n_i}$ as observables and approximate their distribution function by a normal distribution with mean $\sqrt{\lambda_{i}}$ and dispersion $1/4$. This approximation is not very precise (the bias is about 20\,\% for $\lambda=1$), but it seems to be applicable for a coarse estimation. Then, for specific values of $b$, $E_{\mathrm{is}}$  $t$, $\FWHM$, $R_{29}$, and $r_{\mathrm{bg}}$ we calculate the set $\sqrt{\lambda_i}$ according (\ref{ee:1}), and fit it by 
a vector function whose components are given by the square root of (\ref{ee:2}) with additional restriction $J_2=0$. Then we calculate the non-centrality parameter 
\begin{equation}
a^2=4 \sum_{i=1}^N \left(\sqrt{f_i(\theta^s)}-\sqrt{\lambda_i} \right)^2.
\label{ee:a5}
\end{equation}

Our short hypothesis (single peak fit) has 4 degrees of freedom whereas the long hypothesis (double peak fit) has 6 degrees of freedom which yields $r=2$. For significance level $\alpha=0.01$, conditions (\ref{ee:a3}) and (\ref{ee:a4}) are fulfilled at $\lambda_{0.01}(\chi_2^2)=2 \log(100) \simeq 9.21$, and $a_2^2(0.01)\simeq 27.4$. The values of $R_{29}$ necessary to attain $a^2=27.4$ for given values of the other parameters were calculated numerically and presented in Figures~\ref{fig:f4} and \ref{fig:f5}.

\section*{Appendix B. The role of Compton scattering in the formation of an asymmetry or low-energy tail in absorption lines}
In ref.~\cite{Beck07} the 29.18\,keV line shows a low-energy tail, causing a noticeable step of about 10 counts per 3\,eV bin (see Figure~2 (a) in \cite{Beck07}). This step can be caused either by the absorbtion of photons which had 29.18\,keV energy initially, but lost some fraction of it in the source or filter material, or by the escape of some energy fraction from the absorber. The first scenario could be explained by the Compton ``almost forward'' scattering, when a photon loses tiny parts of its energy, about a few eV. The Klein-Nishina differential cross-section of the Compton scattering into the elementary solid angle is
\begin{equation}
\frac{d\sigma}{d\Omega}=\frac{r_e^2}{2}\left(\frac{E_{\gamma}^\prime}{E_\gamma}\right)^2
\left[\frac{E_{\gamma^\prime}}{E_\gamma} +\frac{E_\gamma}{E_{\gamma}^\prime}-\sin^2\theta \right],
\label{eq:A1}
\end{equation}
where $E_{\gamma}$ and $E_{\gamma}^\prime$ are the energies of the photon before and after the Compton scattering respectively, $r_e=e^2/(m_ec^2)$ is the classical electronic radius, and $\theta$ is the scattering angle. Using a well-known relation between $E_{\gamma}^\prime$ and $\theta$, it is easy to express the differential cross-section in units of $E_{\gamma}^\prime$:
\begin{equation}
\frac{d\sigma}{dE_{\gamma}^\prime}=\frac{\pi r_e^2 m_ec^2}{E_\gamma^2}\left(\frac{E_{\gamma}^\prime}{E_\gamma}\right)^2
\left[\frac{E_{\gamma^\prime}}{E_\gamma} +\frac{E_\gamma}{E_{\gamma}^\prime}-\sin^2\theta \right].
\label{eq:A2}
\end{equation}
Note that $E_{\gamma}^\prime=E_{\gamma}-20$\,eV corresponds to $\theta\simeq 8.8^\circ$. For such small $\theta$, (\ref{eq:A2}) yields approximately
\begin{equation}
\frac{d\sigma}{dE_{\gamma}^\prime}\simeq\frac{2 \pi}{E_\gamma^2}r_e^2 m_ec^2 \simeq 3\cdot 10^{-28} \mathrm{cm^2/eV}.
\label{eq:A3}
\end{equation}
The differential probability for the 29.18\,keV photon to be scattered into the 1\,eV range can be coarsely estimated as
\begin{equation}
\frac{dP}{dE_{\gamma}^\prime}=\frac{d\sigma}{dE_{\gamma}^\prime}
\left[\frac{z_\mathrm{U} n_\mathrm{U} \ell_\mathrm{U}}{2}+z_\mathrm{f} n_\mathrm{f} \ell_\mathrm{f} \right],
\label{eq:A4}
\end{equation}
where $z_i$, $n_i$, and $\ell_i$ are the nuclear charge number, the atomic density, and the thickness of the Uranium ($i=\mathrm{U}$) or filter ($i=\mathrm{f}$) material respectively. 

In the experiment~\cite{Beck07}, the Uranium activity of one 19\,mm diameter planchet was about 0.02\,mCi which corresponds to $\ell_\mathrm{U}\simeq 0.3$\,$\mu$m. It was covered by a Titanium foil (``filter'') with $\ell_\mathrm{f}=50.8$\,$\mu$m. An estimation following (\ref{eq:A4}) yields $dP/dE_{\gamma}^\prime \simeq 1.9\cdot 10^{-6}$\,eV$^{-1}$. The peak corresponding to the absorption of 29.18\,keV photons has Gaussian shape with about $3\cdot 10^3$ events per 3\,eV bin height in the maximum, and 26\,eV FWHM, see Figure~2(a) in~\cite{Beck07}. This corresponds to a total number of counts forming this peak of about $3\cdot10^{4}$. Therefore, Compton scattering of 29.18\,keV photons in the source or filter material produces only 0.053 events per 1\,eV interval, or 0.16 events per 3\,eV bin. This value is significantly lower than the observed step of 10 counts per 3\,eV bin. A similar estimation for a $\ell_\mathrm{U}=15\,\mu$m layer of Uranium and $\ell_\mathrm{f}=1.3$\,mm of Aluminium filter (parameters of the planned experiment) leads to $dP/dE_\gamma^\prime\simeq 3.16 \cdot 10^{-5}$\,eV$^{-1}$. 

Note that in (\ref{eq:A4}) we treat all electrons as free. A more accurate estimation requires the substitution of the {\em incoherent scattering function} instead of $z$ but this function does not exceed $z$, see~\cite{Hubbel75} for details. Therefore, a more accurate calculations can only decrease the contribution of Compton scattering to the step of the background count rate.



\end{document}